\documentclass[aps,preprint,onecolumn,superscriptaddress,showpacs,showkeys,nofootinbib]{revtex4-2}
\usepackage{graphicx}
\usepackage{bm}
\usepackage{epsfig}
\usepackage{amsfonts}
\usepackage{amssymb,amscd}
\usepackage{subfigure}
\usepackage{xcolor}
\usepackage{comment}
\usepackage{amsmath}
\usepackage{graphicx}
\usepackage{dcolumn}
\usepackage{bm}
\usepackage{hyperref}

\def\pom{{I\!\!P}}

\begin{document}


\title{Inclusive and diffractive D meson photoproduction in $pPb$ collisions at the LHC using \textsc{Pythia}~8  framework}

\author{Victor P. {\sc Gon\c{c}alves}}
\email{barros@ufpel.edu.br}
\affiliation{Institute of Physics and Mathematics, Federal University of Pelotas, Postal Code 354,  96010-900, Pelotas, RS, Brazil}

\author{Luana {\sc Santana}}
\email{luanas1899@gmail.com}
\affiliation{Institute of Physics and Mathematics, Federal University of Pelotas, Postal Code 354,  96010-900, Pelotas, RS, Brazil}
\affiliation{The Henryk Niewodniczański Institute of Nuclear Physics
Polish Academy of Sciences (PAN), 31-342 Kraków, Poland.}

\author{Wolfgang {\sc Schäfer}}
\email{Wolfgang.Schafer@ifj.edu.pl}
\affiliation{The Henryk Niewodniczański Institute of Nuclear Physics
Polish Academy of Sciences (PAN), 31-342 Kraków, Poland.}

\date{\today}

\begin{abstract}

We present a systematic analysis of inclusive and diffractive $D$-meson photoproduction in $ep$ and ultraperipheral $pPb$ collisions within the \textsc{Pythia}~8 framework. Predictions obtained with different proton and Pomeron PDFs are compared with H1 measurements of inclusive and diffractive $D^*$ photoproduction at HERA. We also present predictions for the rapidity and transverse-momentum distributions of inclusive and diffractive $D^0$ photoproduction at the LHC. Our analysis indicates that the diffractive contribution exceeds 15\% at low $p_T$ and forward rapidities, suggesting that diffractive $D^0$ photoproduction could be accessible in future LHC runs. In addition, measurements of inclusive and diffractive $D^0$ photoproduction in ultraperipheral $pPb$ collisions can help constrain the proton and Pomeron PDFs.

\end{abstract}

\maketitle

\newpage

\section{Introduction}
One of the most promising ways to probe hadron structure is through photon-induced interactions, such as those occurring in deep-inelastic scattering (DIS) and in ultraperipheral hadronic collisions, which are expected to improve our understanding of strong interactions in the high-energy regime\cite{upc}. Current data indicate that the gluon density inside the proton increases with energy and that, in this regime, the hadron becomes a dense system in which nonlinear effects inherent to QCD dynamics may become visible. Although our understanding of QCD at high energies has advanced substantially in recent years, several open questions remain~\cite{Morreale:2021pnn}. Consequently, the assumptions underlying different approaches must still be tested by comparing their predictions with future high-energy experimental data.

A prominent testing ground for different QCD approaches is $D$-meson production in photon-induced interactions~\cite{Klein:2000dk,Klein:2002wm,Goncalves:2003is,Goncalves:2004dn,Goncalves:2009ey,Goncalves:2013oga,Goncalves:2015cik,Kotko:2017oxg,Goncalves:2017zdx,Gimeno-Estivill:2025rbw,Goncalves:2025wwt,Cacciari:2025tgr,Goncalves:2026lea,Guzey:2026lfb}. The associated cross section can be computed perturbatively and is sensitive to the target gluon distribution at small values of the Bjorken-$x$ variable. Recently, this final state has been measured in ultraperipheral $PbPb$ collisions at the LHC~\cite{CMS:2025jjx,Nese:2025ohz}, and more precise data are expected in the coming years, opening the possibility of using this process to constrain QCD dynamics at high energies. This prospect has motivated several groups to refine the theoretical description of inclusive $D$-meson photoproduction~\cite{Gimeno-Estivill:2025rbw,Goncalves:2025wwt,Cacciari:2025tgr,Goncalves:2026lea}, where the target hadron breaks up, and to study diffractive $D$-meson photoproduction in ultraperipheral hadronic collisions~\cite{Guzey:2026lfb}, where the target remains intact. In particular, Ref.~\cite{Gimeno-Estivill:2025rbw} works within the Color Glass Condensate framework, while Refs.~\cite{Goncalves:2025wwt,Goncalves:2026lea} estimate the cross section in the color-dipole $S$-matrix formalism. Despite the different assumptions, both approaches yield similar expressions for the heavy-quark photoproduction cross section in terms of unintegrated gluon distributions. By contrast, Refs.~\cite{Cacciari:2025tgr,Guzey:2026lfb} compute the photoproduction cross section assuming the collinear factorization (for inclusive and diffractive channels) and within the generalized photon--nucleus fixed-order next-to-leading-logarithm framework, in which the cross section is written in terms of integrated parton distributions evolved with DGLAP.
{ The diffractive contribution has also been estimated at the level of heavy quarks, without fragmentation to $D$-mesons, in Ref.~\cite{Goncalves:2006xi}   using color-dipole approach, and discussed in more detail in Refs.~\cite{ReinkePelicer:2018gyh,Goncalves:2019jdl,Linek:2023kga}. In particular, Refs.\cite{ReinkePelicer:2018gyh,Linek:2023kga} have treated this process in terms of generalized transverse momentum dependent parton distribution.}

\begin{figure}
    \centering
        \includegraphics[width=0.65\linewidth]{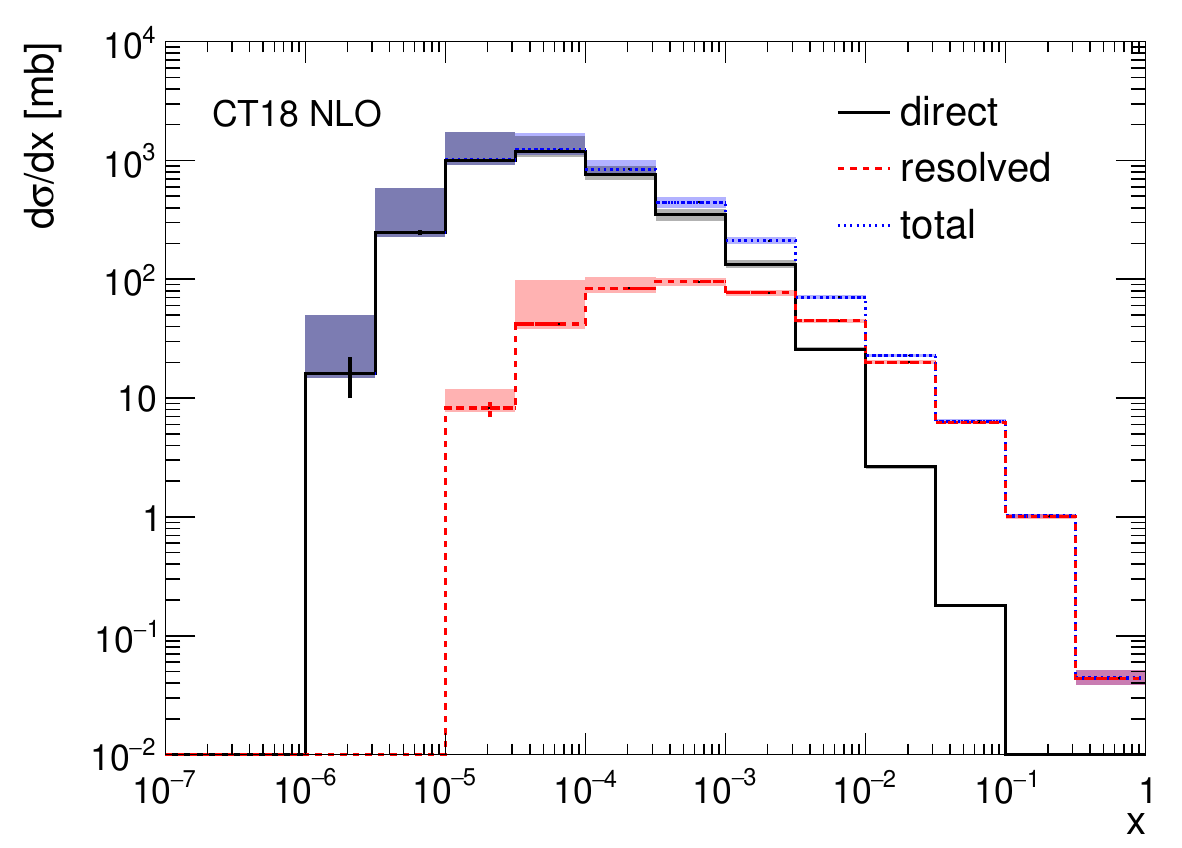}
    \caption{ Differential distribution for the inclusive $D^0$ meson photoproduction in ultraperipheral $pPb$ collisions at $\sqrt{s} = 8.1$ TeV as a function of the Bjorken  $x$ variable. }
    \label{fig:map}
\end{figure}

Our goal in this paper is to contribute to these recent efforts by presenting predictions for inclusive and diffractive $D$-meson photoproduction in ultraperipheral hadronic collisions at the LHC, obtained within the \textsc{Pythia}~8 framework~\cite{Bierlich:2022pfr}. In recent years, the description of photon-induced interactions in this Monte Carlo generator has been improved and extended to ultraperipheral collisions~\cite{Friberg:2000ra,Helenius:2017aqz,Helenius:2019gbd,Helenius:2024vdj}. The current version also allows us to estimate heavy-meson photoproduction cross sections in $ep$ collisions and to compare its predictions with existing HERA $ep$ data on $D^*$-meson production in inclusive and diffractive channels. We focus on $D^0$ photoproduction in ultraperipheral $pPb$ collisions, which are dominated by photon--proton interactions, with the nucleus acting as the photon emitter.  { As illustrated in Fig.~\ref{fig:map}, the differential distribution in the Bjorken $x$ variable for the inclusive production shows that, at LHC energies, both direct and resolved photon processes are dominated by the small-$x$ regime. This region can be probed in greater detail by selecting forward events. Consequently, $D$ photoproduction at the LHC is expected to provide valuable constraints on the proton structure at small $x$.}

We estimate inclusive and diffractive $D$-meson cross sections using different parameterizations of the proton and pomeron parton distribution functions (PDFs), respectively, and including both direct and resolved photon contributions. We compare with HERA $ep$ data and present predictions for transverse-momentum and rapidity distributions in ultraperipheral $pPb$ collisions at $\sqrt{s}=8.1\,\mathrm{TeV}$. Additionally, we estimate the ratio of diffractive to inclusive $D$-meson photoproduction cross sections for different proton PDFs. Finally, we also compute the contribution from  $D$-meson production via $\gamma\gamma$ interactions.

This paper is organized as follows. In the next section, we describe the \textsc{Pythia}~8 setup used to simulate inclusive and diffractive $D$-meson photoproduction in $ep$ and ultraperipheral $pPb$ collisions, and we discuss the proton and pomeron PDFs employed in our calculations. In Section~\ref{sec:results}, we compare the \textsc{Pythia}~8 predictions with HERA data on inclusive and diffractive $D^*$ production and present our predictions for ultraperipheral collisions. In particular, we consider $pPb$ collisions at $\sqrt{s}=8.1\,\mathrm{TeV}$ and present transverse-momentum and rapidity distributions for $D^0$ photoproduction in inclusive and diffractive channels. We also report the ratio of diffractive to inclusive cross sections and the contribution of $\gamma\gamma$ interactions to $D^0$-meson production. Finally, in Section~\ref{sec:summary}, we summarize our main results and conclusions.

\section{Simulation setup}
In our analysis, we use the \textsc{Pythia}~8 event generator (version 8.317) to simulate inclusive and diffractive $D$-meson photoproduction in $ep$ and ultraperipheral $pPb$ collisions. The generator models the hard scattering of two particles from the incoming beams into outgoing partons, followed by QCD radiation (gluon emission and $q\bar q$ splittings) prior to hadronization. This evolution is described by a parton-shower algorithm based on DGLAP evolution. 
{Finally, the produced heavy quarks hadronized using the Lund string model, where they act as  string endpoints and are converted into  color-neutral heavy hadrons during string breaking.}

In the case of photoproduction processes, the flux of photons associated with the incoming charged particle ($e$ or $Pb$) can be factorized from the hard-process cross section, and  \textsc{Pythia}~8 simulates both the direct and resolved - photon contributions~\cite{Helenius:2019gbd}. Additionally, hard diffractive events in photoproduction are modeled using the Ingelman–Schlein approach~\cite{Ingelman:1984ns} with a dynamical gap survival based on the framework for multiparton interactions~\cite{Rasmussen:2015qgr}. It is important to emphasize that the dynamical gap survival is present only in the resolved - photon contribution.  
In the collinear-factorization approach implemented in \textsc{Pythia}~8, the hard partonic subprocesses are the same for inclusive and diffractive heavy-quark photoproduction and are dominated by photon--gluon fusion ($\gamma g \to Q\bar Q$).

The key ingredients in our photoproduction cross-section estimates are the photon fluxes of the incoming particles and the inclusive/diffractive parton distribution functions of the proton target. Following Ref.~\cite{Helenius:2019gbd}, we model the electron photon flux using the Weizs\"acker--Williams approximation~\cite{Budnev:1975poe}, while the nuclear photon flux is taken as~\cite{upc}
\begin{eqnarray}
    f_{\gamma}^A(x) = \frac{2 Z^2 \alpha_{\mathrm{em}}}{\pi}\,\frac{1}{x}\left[\xi K_0(\xi)K_1(\xi) - \frac{\xi^2}{2}\left(K_1^2(\xi) - K_0^2(\xi)\right)\right] \,\,,
\end{eqnarray}
where $x$ is the photon energy fraction, $\alpha_{\mathrm{em}}$ is the electromagnetic fine-structure constant, and $Z$ is the nuclear charge. We define $\xi = x m b_{\min}$, with $m$ the per-nucleon mass and $b_{\min}$ the minimum impact parameter, taken as the sum of the radii of the incoming particles. For inclusive interactions, we consider four proton-PDF parameterizations: CT18NLO \cite{Hou:2019CT18}, MSHT20NLO \cite{Cridge:2021MSHT}, NNPDF40NLO \cite{Ball:2022NNPDF}, and DGHP (Parameterization 1)~\cite{Duwentaster:2023mbk}. These sets yield different small-$x$ gluon distributions; in particular, DGHP includes leading nonlinear corrections to DGLAP evolution.

For diffractive interactions, we follow the standard Ingelman--Schlein approach~\cite{Ingelman:1984ns} and model the diffractive parton distribution functions (dPDFs) in terms of a Pomeron flux emitted by the proton and parton densities inside the Pomeron. In this framework, the diffractive gluon distribution in the proton, $f_g^D(x,\mu^2)$, is written as a convolution of the proton--Pomeron flux factor $f_{I\!\!P/p}(x_{\pom})$---with $x_{\pom}$ the longitudinal momentum fraction of the incoming hadron carried by the Pomeron---and the gluon density in the Pomeron, $f_{g/I\!\!P}(\beta,\mu^2)$, where $\beta$ is the momentum fraction of the struck parton inside the Pomeron. The flux factor is obtained by integrating over the squared momentum transfer $t$,
\begin{eqnarray}
  f_{I\!\!P/p}(x_{\pom}) &=& \int_{t_{\min}}^{t_{\max}} \! dt\; f_{\pom/p}(x_{\pom},t) \,\,,
\end{eqnarray}
with
\begin{eqnarray}
  f_{\pom/p}(x_{\pom},t) &=& A_{\pom}\,\frac{e^{\mathcal{B}_{\pom} t}}{x_{\pom}^{2\alpha_{\pom}(t)-1}} \,\,,
\end{eqnarray}
where $t_{\max}\simeq -\frac{m_p^2 x_{\pom}^2}{1-x_{\pom}}$ is the maximum kinematically allowed value and $t_{\min}=-1.0\,\mathrm{GeV}^2$ corresponds to the experimental acceptance. The flux is motivated by Regge theory with a linear Pomeron trajectory, $\alpha_{\pom}(t)=\alpha_{\pom}(0)+\alpha_{\pom}^\prime t$, and the parameters $\mathcal{B}_{\pom}$ and $\alpha_\pom(0),\alpha_{\pom}^\prime$ (with uncertainties) are taken from fits to HERA data. The resulting diffractive gluon distribution reads
\begin{eqnarray}
  f_g^D(x,\mu^2) &=& \int \frac{dx_{\pom}}{x_{\pom}}\, f_{I\!\!P/p}(x_{\pom})\, g_{I\!\!P}\left(\frac{x}{x_{\pom}},\mu^2\right) \,\,.
\label{Eq:difPDF}
\end{eqnarray}
Analogous expressions hold for the diffractive quark and antiquark distributions. In our analysis, we consider two dPDF sets fitted to the combined HERA data: H1 NLO 2006 (Fit A) \cite{Aktas:2006H1} and GKG18 NLO (Fit A) \cite{Goharipour:2018GKG}. As discussed in Ref.~\cite{Helenius:2019gbd}, they predict similar Pomeron fluxes but differ in the parton densities inside the Pomeron. 

Finally, cross sections for the resolved-photon contribution are computed using the CJKL photon PDF parametrization~\cite{Cornet:2002iy}, the default choice in \textsc{Pythia}~8. We also implement \textsc{SoftQCD:nonDiffractive = on}, which activates the inelastic non-diffractive (ND) component of hadron--hadron interactions, and set \textsc{StringFlav:mesonCvector = 1.5}, which controls charm vector-meson suppression.  As shown in Ref.~\cite{Altmann:2024kwx}, the value 1.5 improves the description of the $D^{*+}/D^0$ ratio measured in $e^+e^-$ collisions at LEP. The configuration card used to derive the results presented in the next section is available from the authors upon request.

\section{Results}
\label{sec:results}

In what follows, we present \textsc{Pythia}~8 predictions for inclusive and diffractive heavy-meson photoproduction in $ep$ and ultraperipheral $pPb$ collisions, using the PDFs discussed in the previous section. We first compare our results with $D^*$ data measured by the H1 Collaboration in $ep$ collisions at HERA~\cite{H1:2011myz,H1:2006zxb} to validate the framework. We then study $D^0$ production in ultraperipheral $pPb$ collisions, keeping the same assumptions for charm photoproduction and changing only the fragmentation function to model hadronization of the charm quark into a $D^0$ meson.

\begin{figure}[t]
    \centering
    \includegraphics[width=0.45\linewidth]{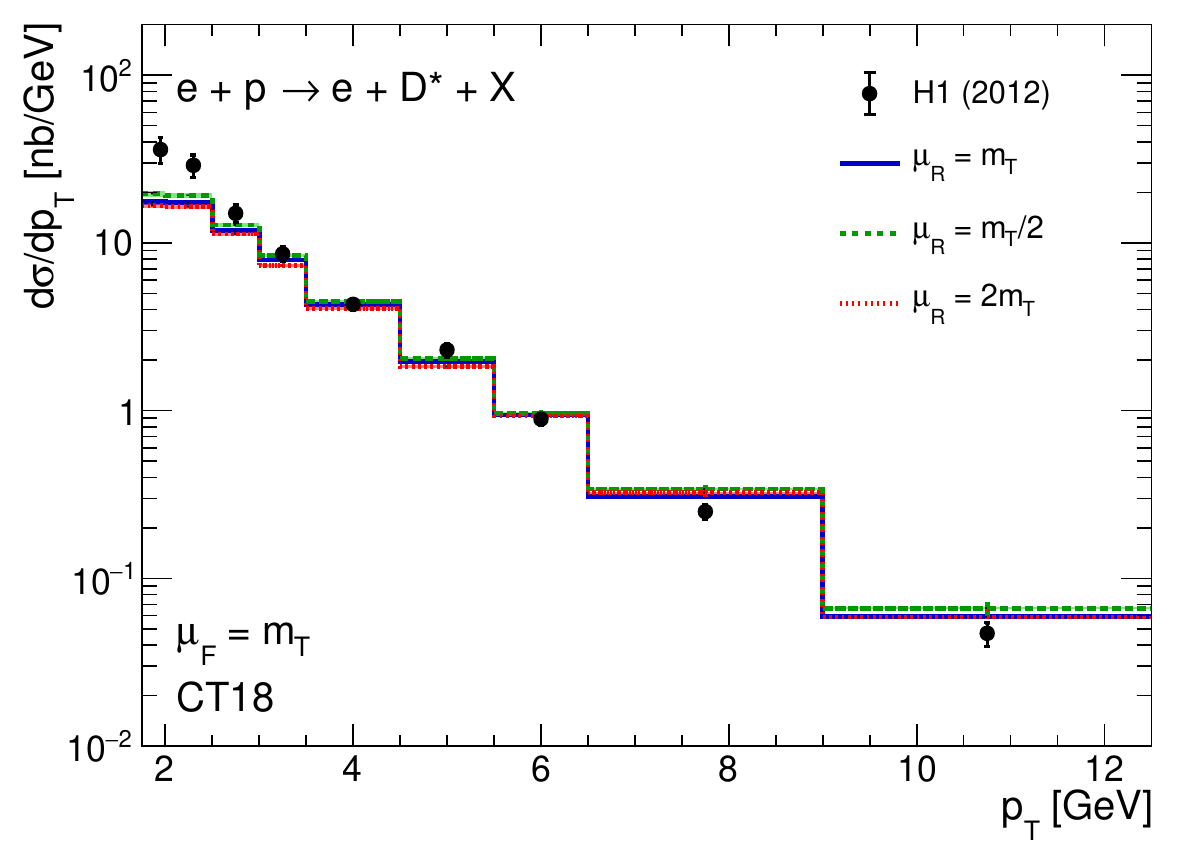}
    \includegraphics[width=0.45\linewidth]{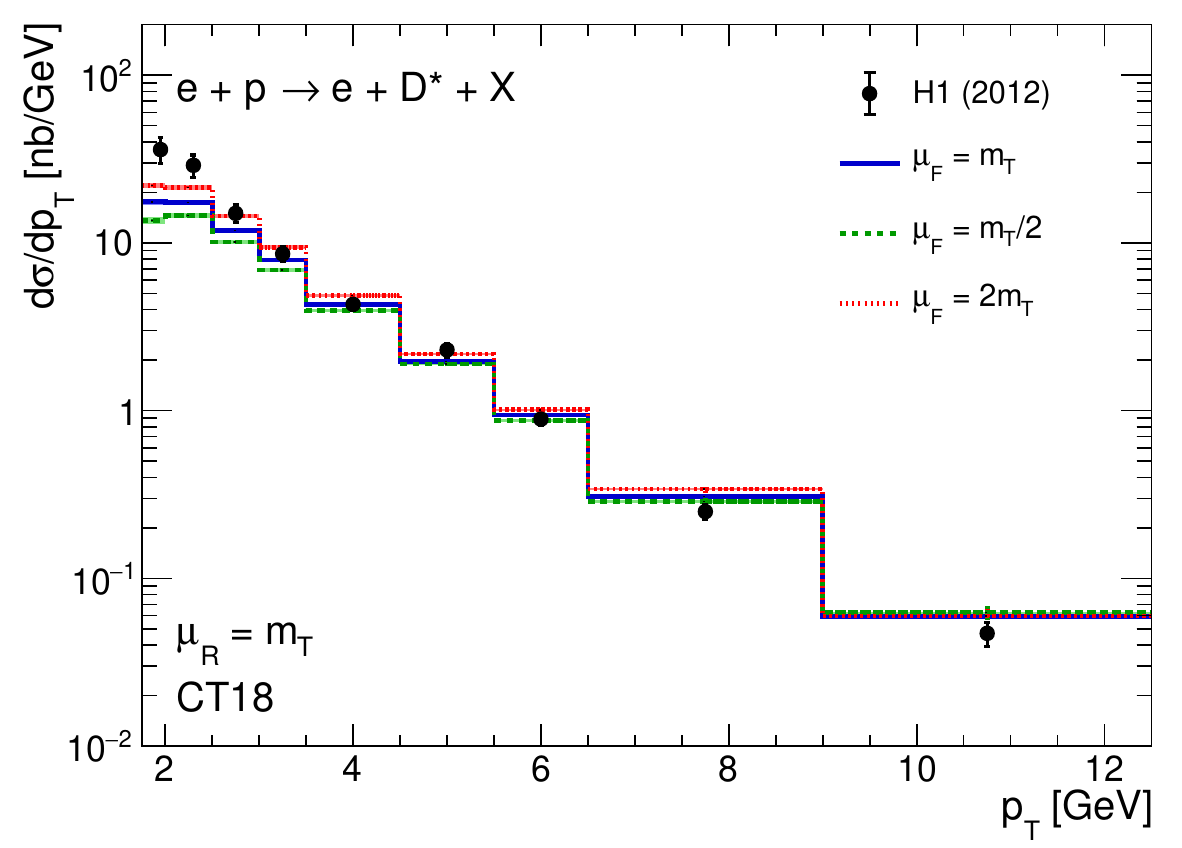}
    \caption{Single differential $D^*$
cross-section as a function of $p_T$ compared to \textsc{Pythia}~8 predictions, derived considering distinct choices for the renormalization (left) and factorization (middle) scales, as well as different inclusive PDFs for the proton (right). Data from H1 Collaboration~\cite{H1:2011myz}.}
    \label{fig:incDdist}
\end{figure}

\subsection{$ep$ collisions}
The H1 Collaboration has measured the inclusive $D^*$ photoproduction cross - section in $ep$ collisions by selecting  events characterized by small photon virtualities $Q^2 < 2$ GeV$^2$, values of the $D^*$ pseudorapidity $|\eta| \le 1.5$, and photon - proton center-of-mass energies in the range  $100 < W_{\gamma p} < 285$ GeV~\cite{H1:2011myz}. These limits are applied into \textsc{Pythia}~8. 

In what follows, we compare \textsc{Pythia}~8 predictions with the measured single-differential cross sections as a function of the $D^*$ transverse momentum ($p_T$), derived using the CT18 proton PDF. Moreover, we set the factorization and renormalization scales equal to the charm transverse mass, i.e. $\mu_R = \mu_F = \mu$, with $\mu = m_T = \sqrt{p_T^2 + m_c^2}$. The dependence on this choice is examined in  Fig.~\ref{fig:incDdist}, where we vary one scale while keeping the other fixed. We find that the low-$p_T$ predictions are sensitive to $\mu_F$ but still underestimate the data, whereas at large $p_T$ the description improves and the scale dependence is mild.

\begin{figure}[t]
    \centering
    \includegraphics[width=0.32\linewidth]{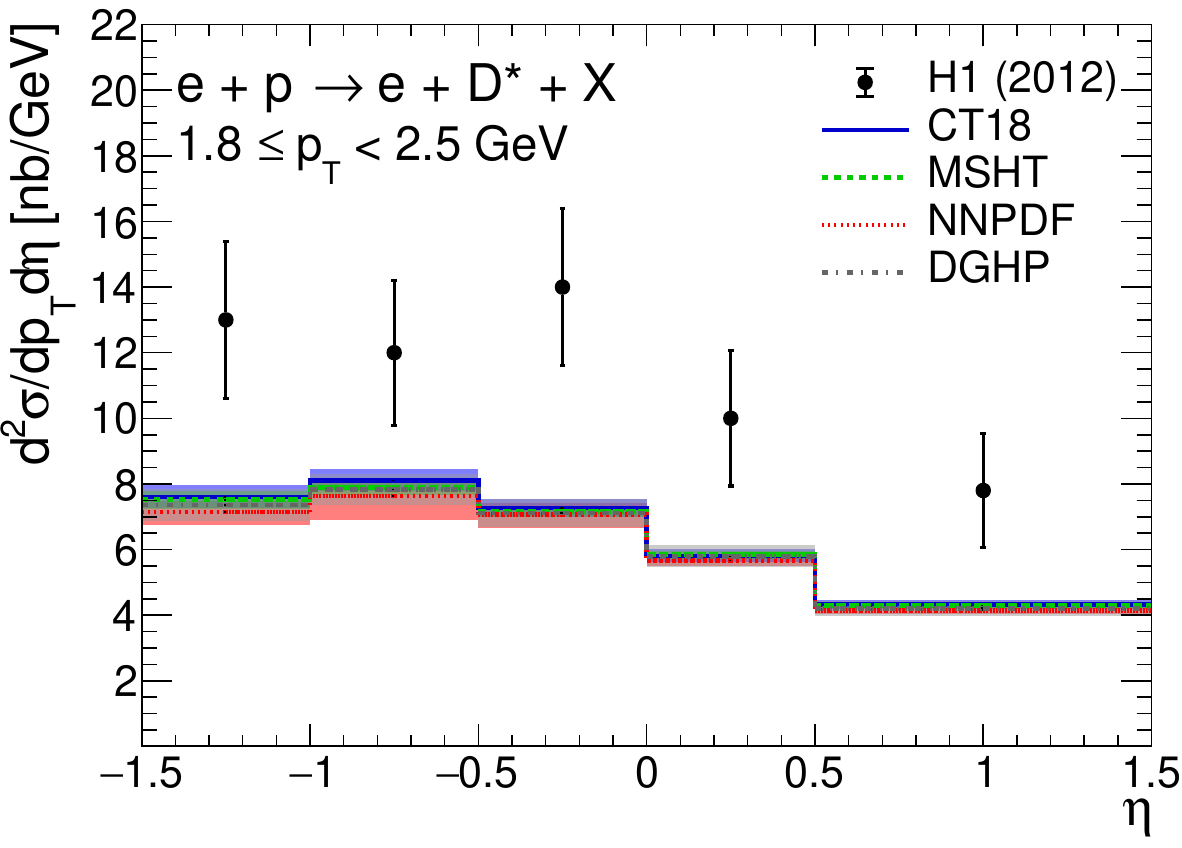}
    \includegraphics[width=0.32\linewidth]{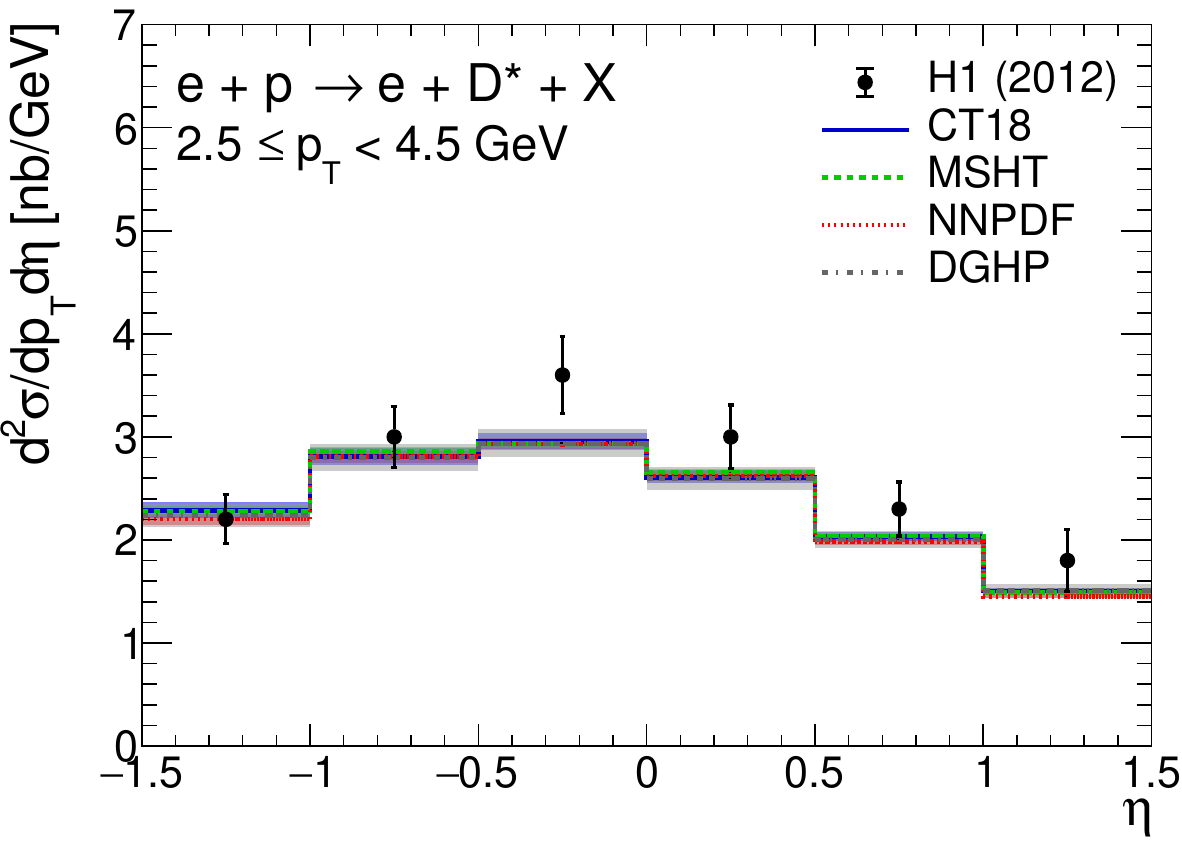}
    \includegraphics[width=0.32\linewidth]{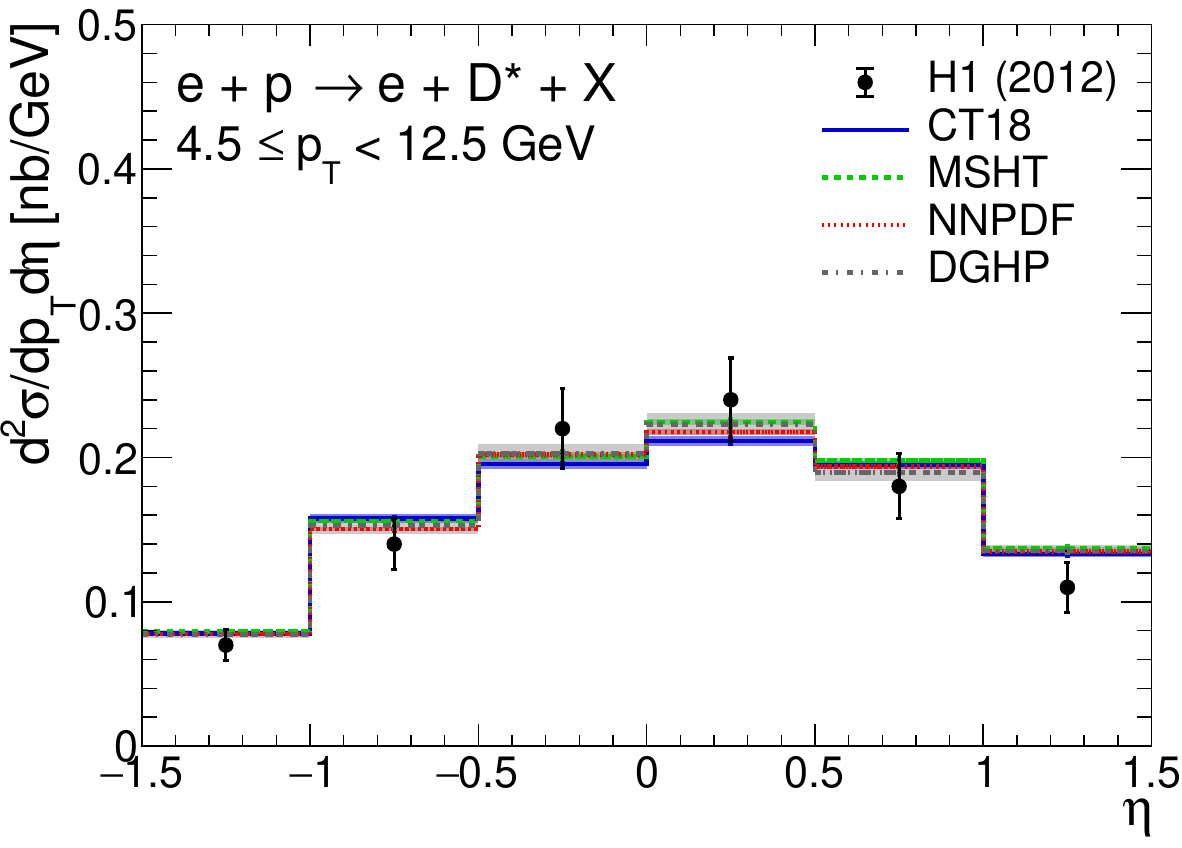}
    \caption{Double differential $D^*$
cross-section as a function of  $\eta$ for
three bins of $p_T$, compared to \textsc{Pythia}~8 predictions, derived considering different inclusive PDFs for the proton. Data from H1 Collaboration~\cite{H1:2011myz}.}
    \label{fig:incDbins}
\end{figure}

Figure~\ref{fig:incDbins} compares the predictions derived assuming different proton PDFs for the double-differential $D^0$ cross sections ($d^2\sigma/d\eta dp_T$) with H1 data across three $p_T$ bins. Alongside predictions obtained with central PDF sets, we show uncertainty bands corresponding to the variations provided by each group. \textsc{Pythia} 8 describes the data well, with the largest tension occurring at low $p_T$, in agreement with the results shown in Fig.~\ref{fig:incDdist}.  These comparisons indicate that the inclusive cross section is sensitive to the choice of proton PDFs, suggesting that future measurements of this final state will help constrain the gluon distribution.

\begin{figure}
    \centering
    \includegraphics[width=0.32\linewidth]{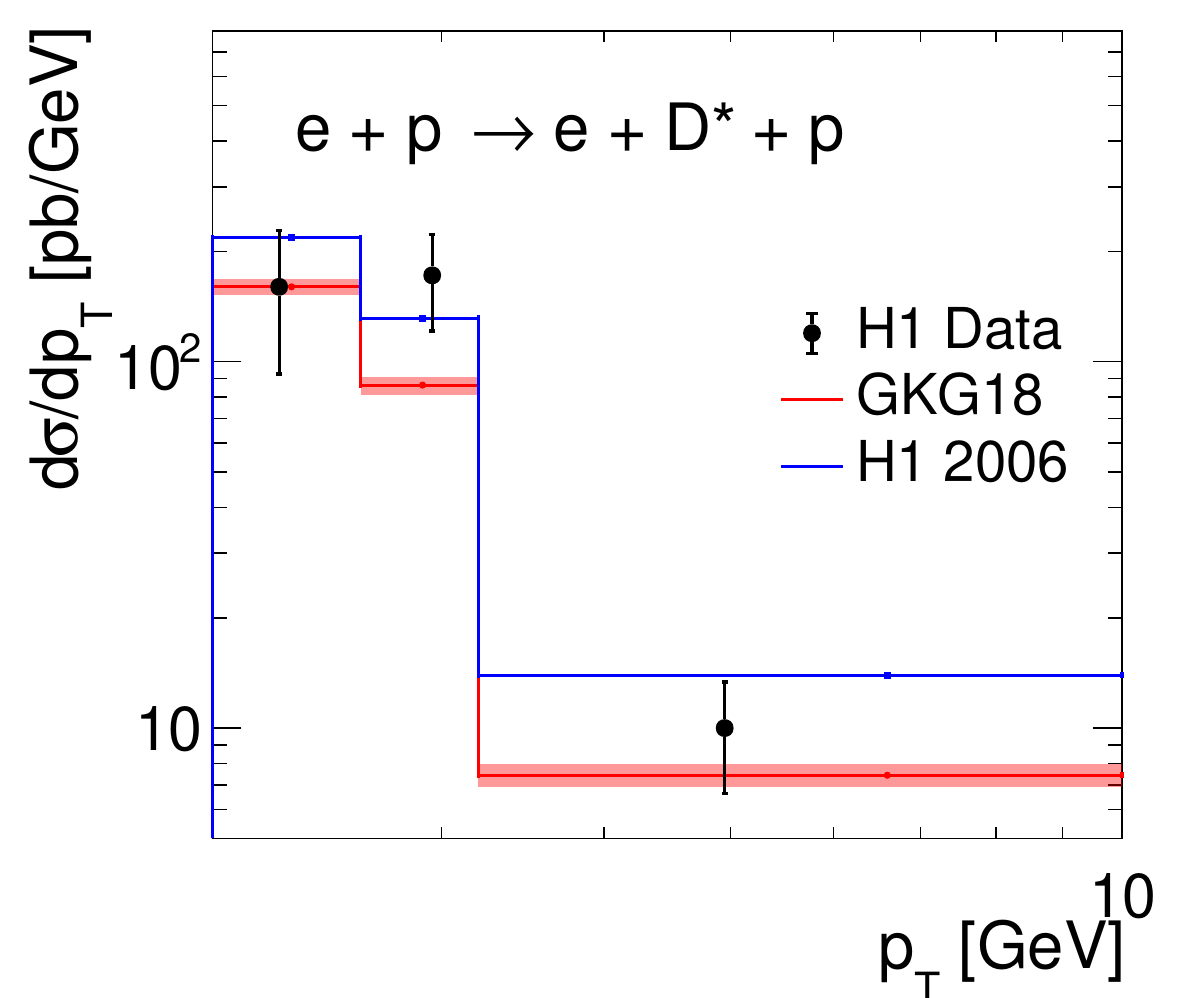}
    \includegraphics[width=0.32\linewidth]{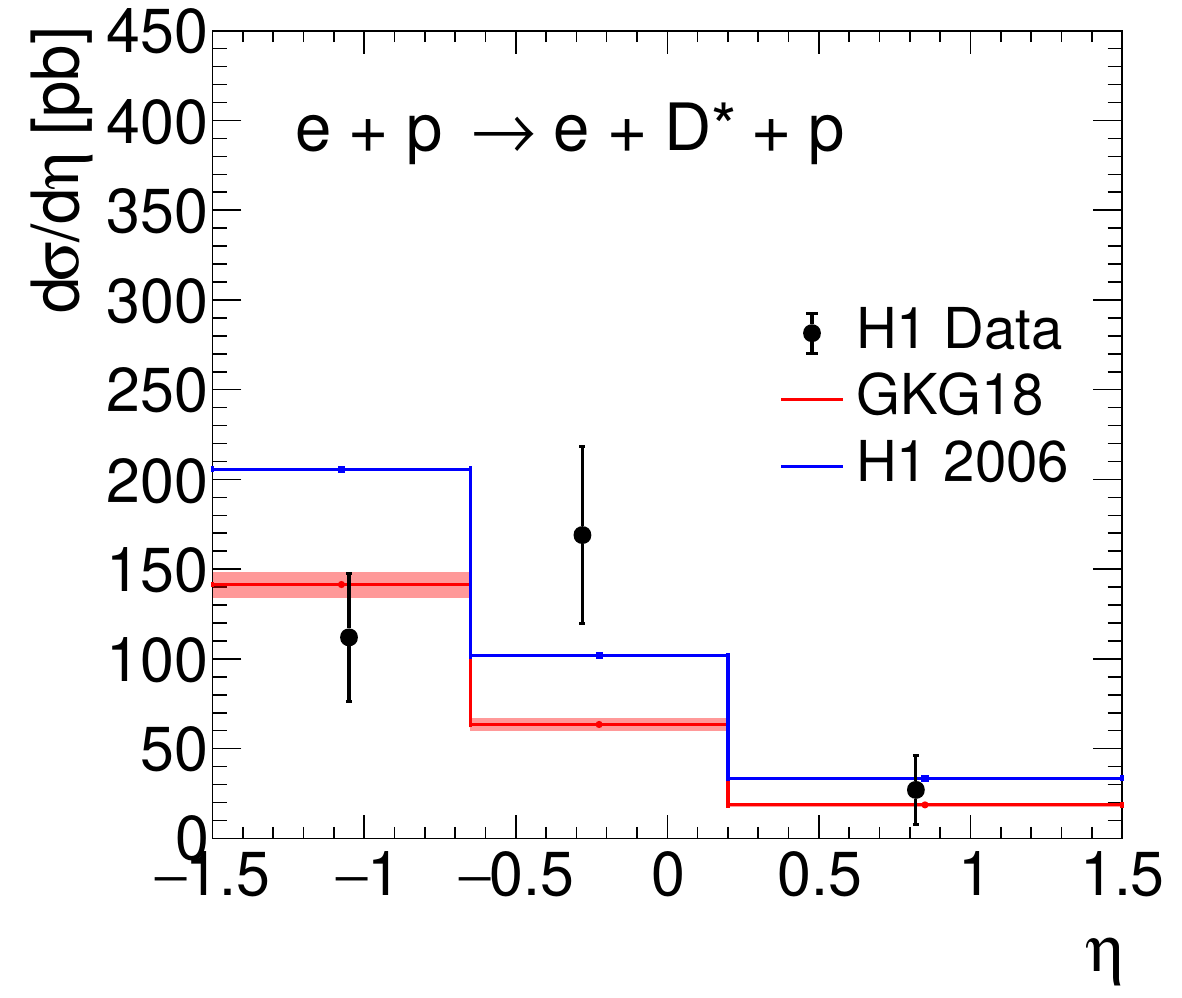}
    \includegraphics[width=0.32\linewidth]{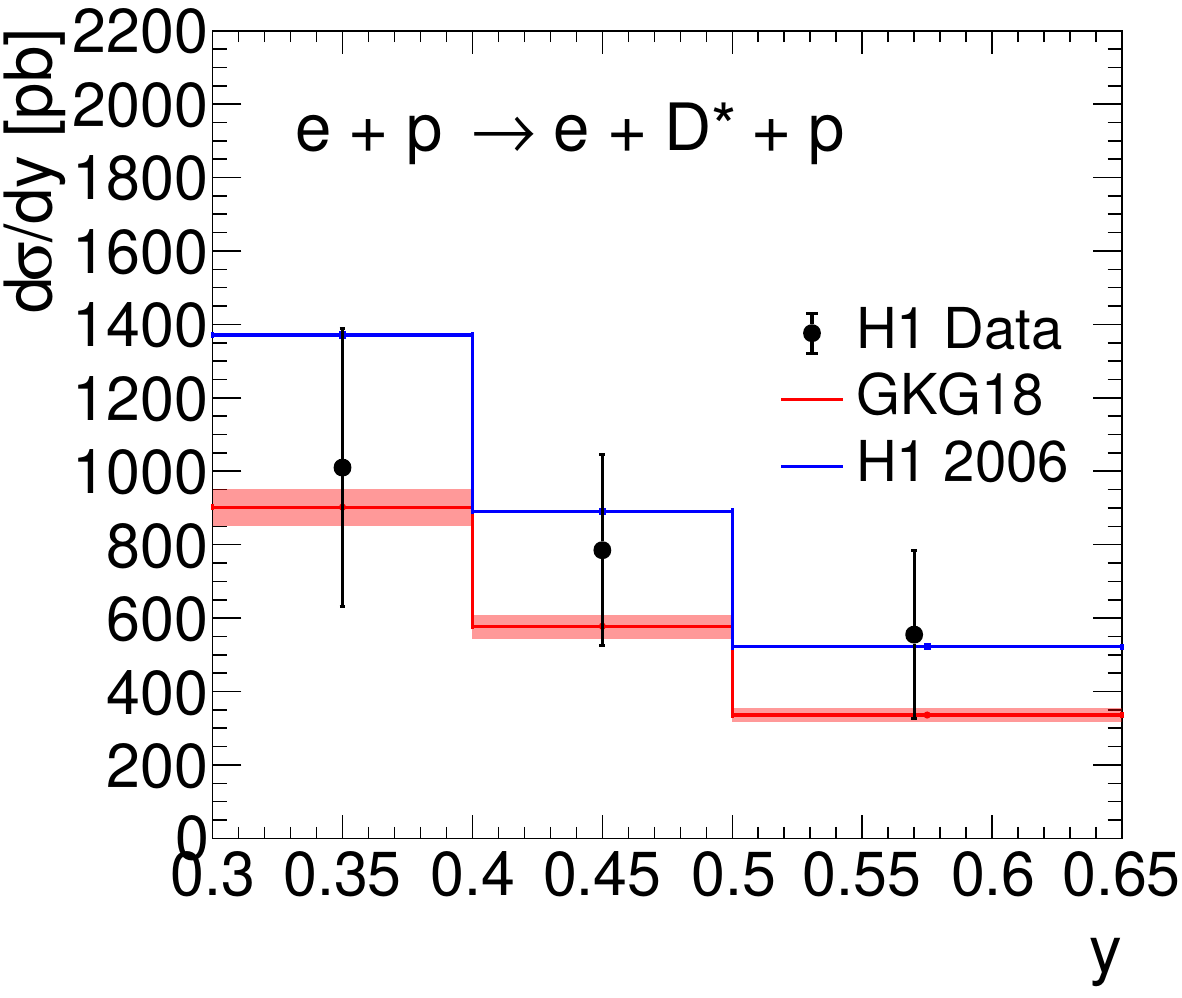}
    \caption{Differential cross-sections for diffractive $D^*$ photoproduction as a function of $p_T$, $\eta$ and the inelasticity $y$. Data from H1 Collaboration~\cite{H1:2006zxb}.}
    \label{fig:difHERA}
\end{figure}

Diffractive $D^*$ photoproduction was also measured by the H1 Collaboration, with the data reported in Ref.~\cite{H1:2006zxb}. Events were selected using the cuts $Q^2 < 0.01$~GeV$^2$, $0.3 < y < 0.65$, $x_{\pom} < 0.04$, $|t| < 1$~GeV$^2$, $p_T > 2$~GeV, and $|\eta| < 1.5$, which we implement in \textsc{Pythia}~8. The resulting predictions, obtained with different diffractive PDF (dPDF) parameterizations, are compared with the data in Fig.~\ref{fig:difHERA}. The predictions are sensitive to the choice of dPDF, with the H1 set yielding larger differential cross sections. However, the limited HERA dataset does not allow one to discriminate between the different parameterizations. These results indicate that \textsc{Pythia}~8 provides a satisfactory baseline for describing diffractive heavy-meson photoproduction.

\subsection{Ultraperipheral $pPb$ collisions}
We consider inclusive and diffractive photoproduction in ultraperipheral $pPb$ collisions at $\sqrt{s}=8.1$~TeV. We take the ion as the photon source and the proton as the target, since the photon flux scales as $Z^2$. We also define the positive-rapidity direction as that of the photon-emitting ion.

\begin{figure}
    \centering
    \includegraphics[width=0.45\linewidth]{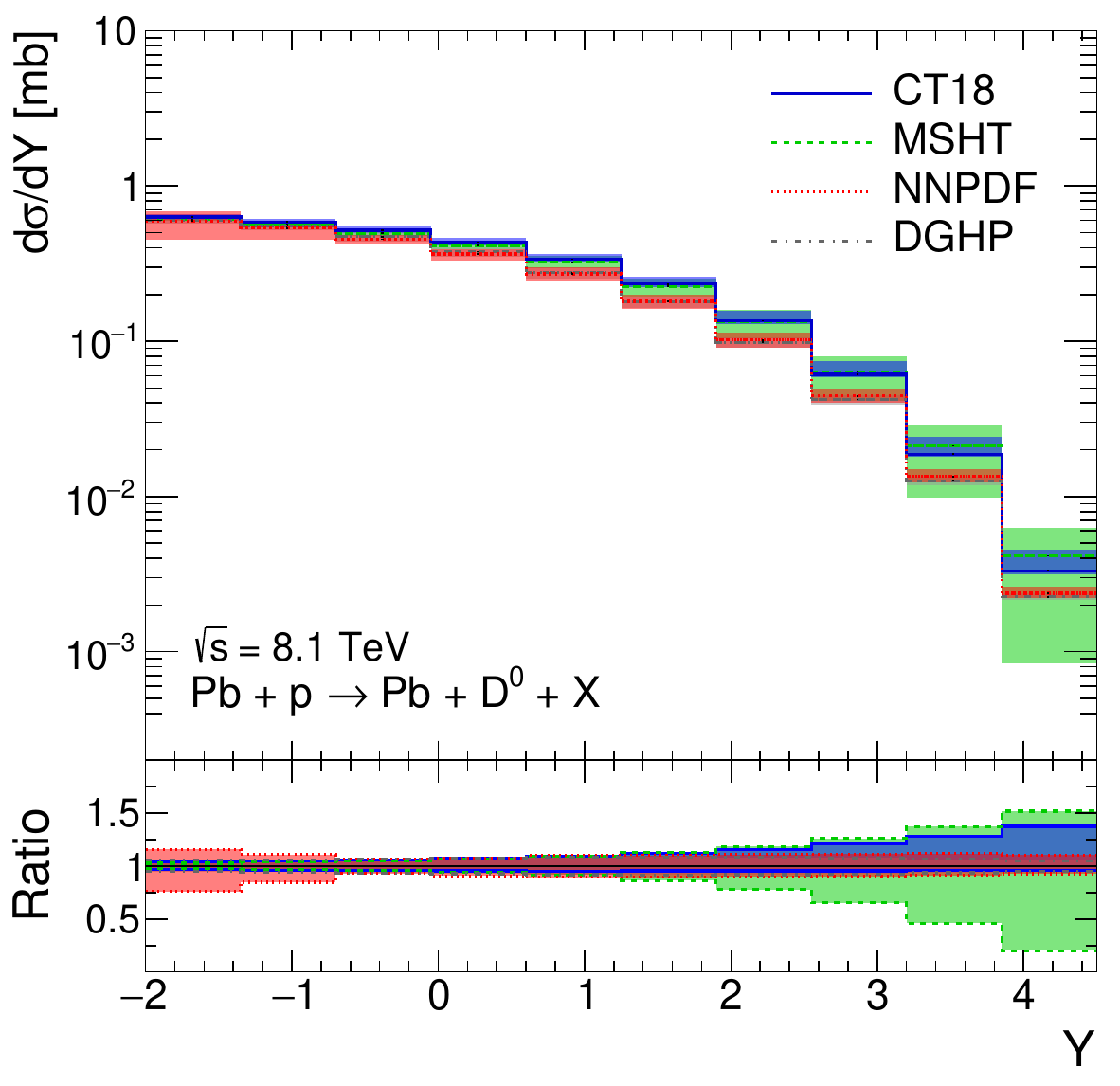}
    \caption{Predictions for the rapidity distribution for the inclusive $D^0$ photoproduction in ultraperipheral $pPb$ collisions, derived using the \textsc{Pythia}~8 framework and assuming distinct proton PDFs.}
    \label{fig:upc_inc_rap}
\end{figure}

\begin{figure}[t]
    \centering
    \includegraphics[width=0.32\linewidth]{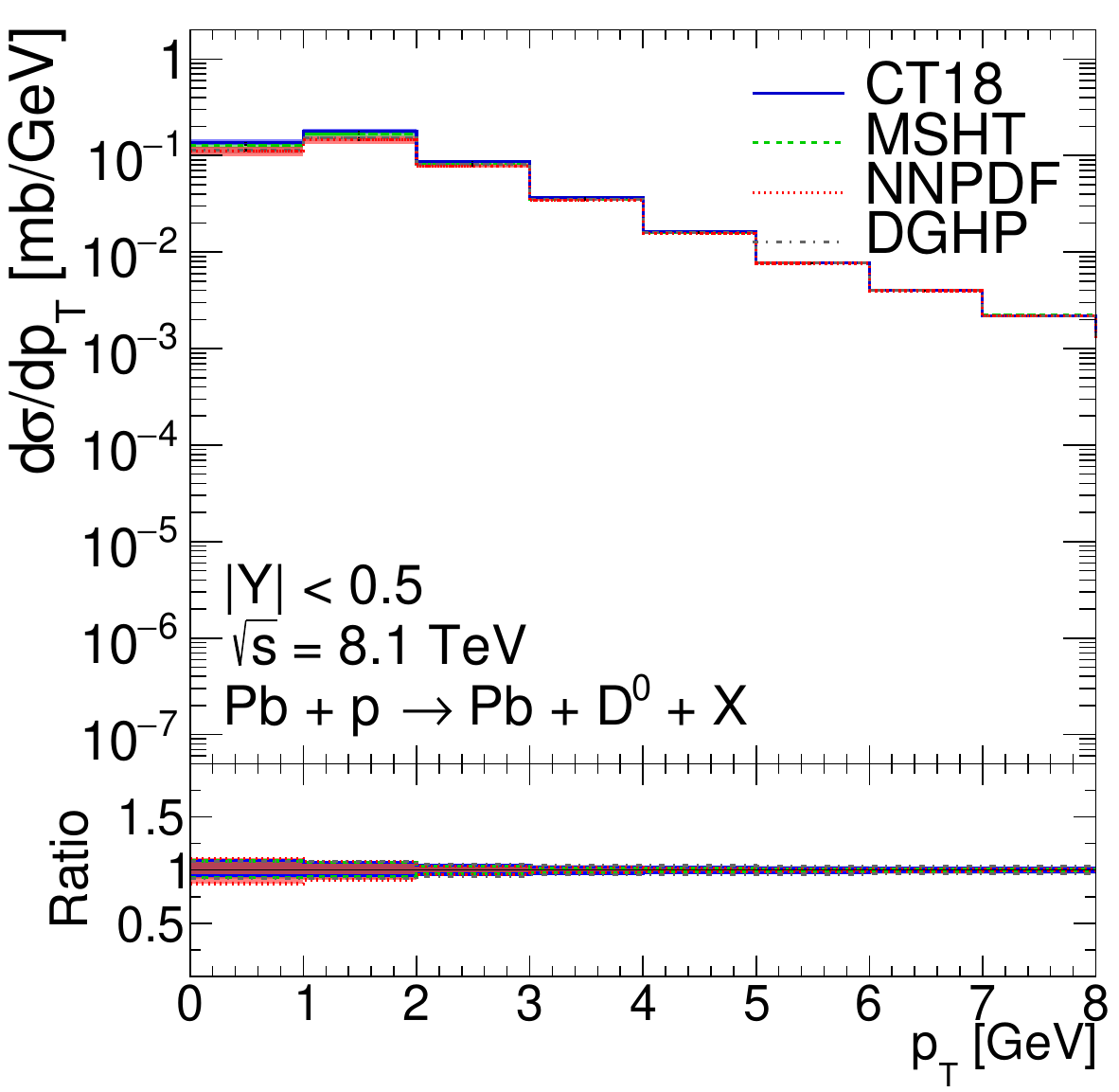}
    \includegraphics[width=0.32\linewidth]{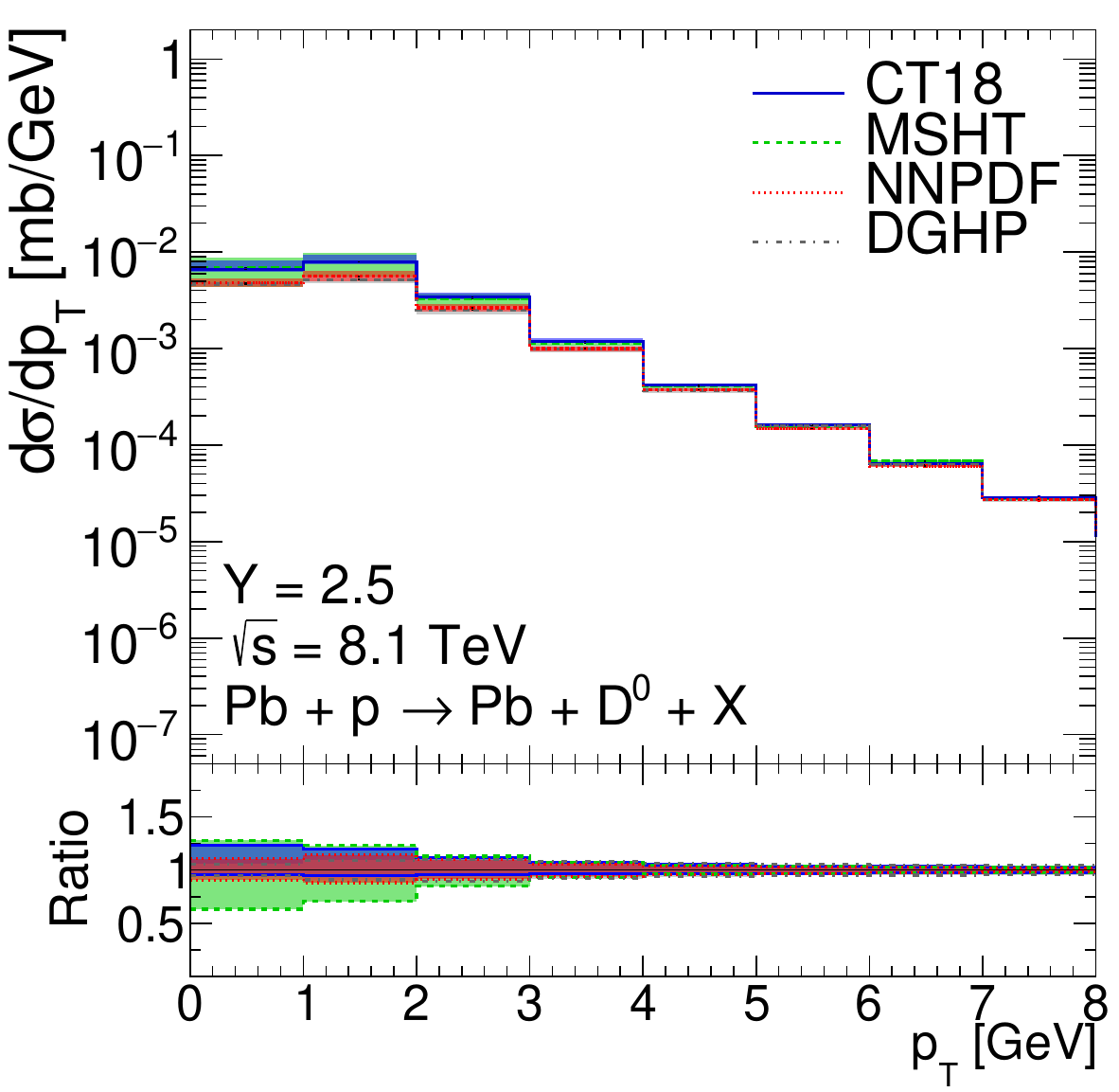}
    \includegraphics[width=0.32\linewidth]{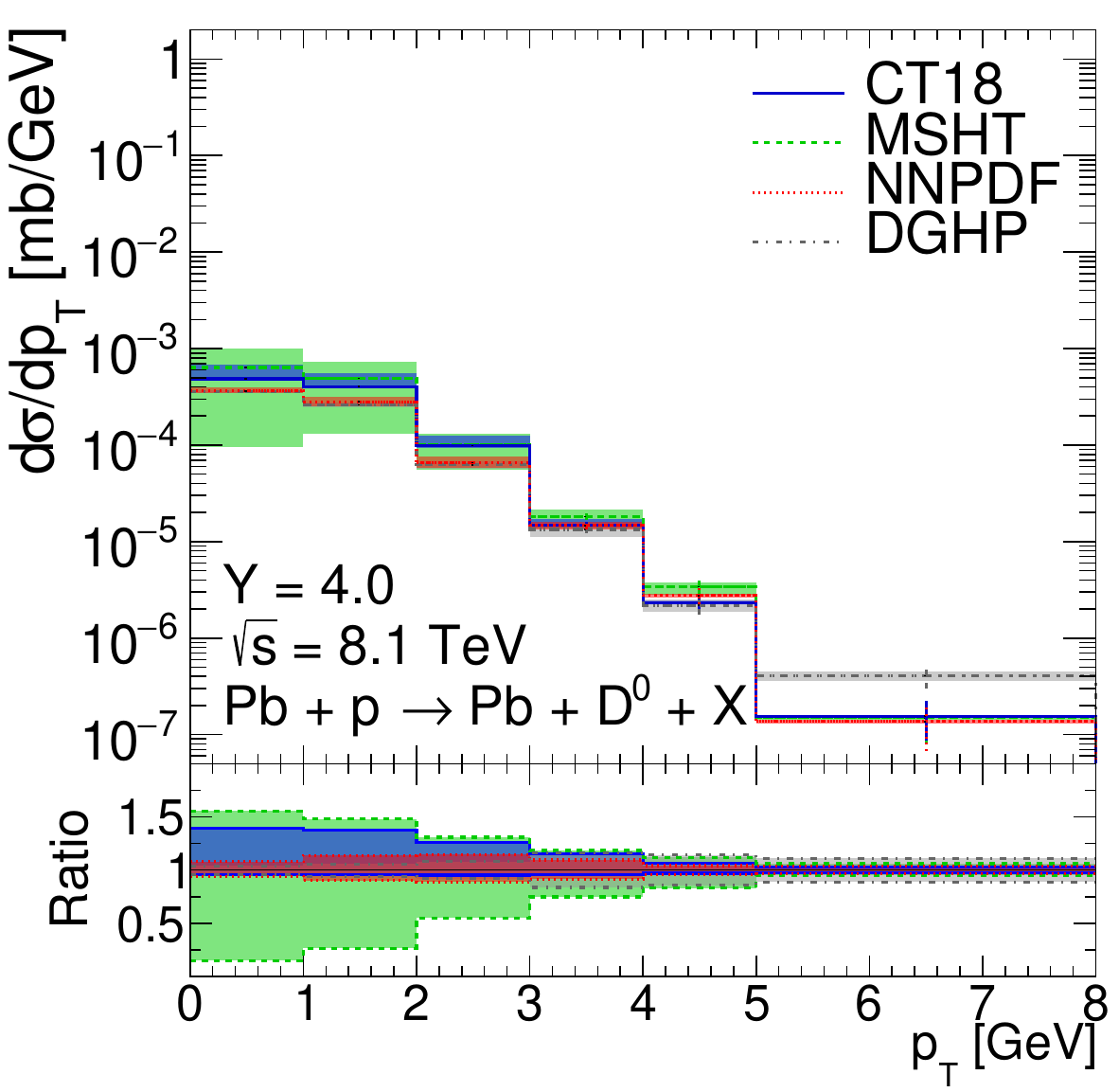}

    \caption{Predictions for the transverse momentum distribution for the inclusive $D^0$ photoproduction at different rapidities  in ultraperipheral $pPb$ collisions, derived using the \textsc{Pythia}~8 framework and assuming distinct proton PDFs.}
    \label{fig:upc_inc_pt}
    \end{figure}

In Fig.~\ref{fig:upc_inc_rap} we present our results for the rapidity distribution for inclusive $D^0$ photoproduction in ultraperipheral $pPb$ collisions, obtained using different proton PDFs. The predictions are similar at small rapidities, but the differences grow with increasing $Y$. This behavior is expected, since larger rapidities probe smaller values of $x$ in the proton, where the gluon distributions from different parameterizations diverge and carry larger uncertainties. The corresponding transverse-momentum distributions at different meson rapidities are shown in Fig.~\ref{fig:upc_inc_pt}. We find that the shapes predicted by the different PDFs are largely similar, with the main differences appearing at larger rapidities and low $p_T$, as highlighted in the lower panels, which show the ratio of the results obtained with the individual replicas to the corresponding central set for each PDF.

\begin{figure}[t]
    \centering
    \includegraphics[width=0.49\linewidth]{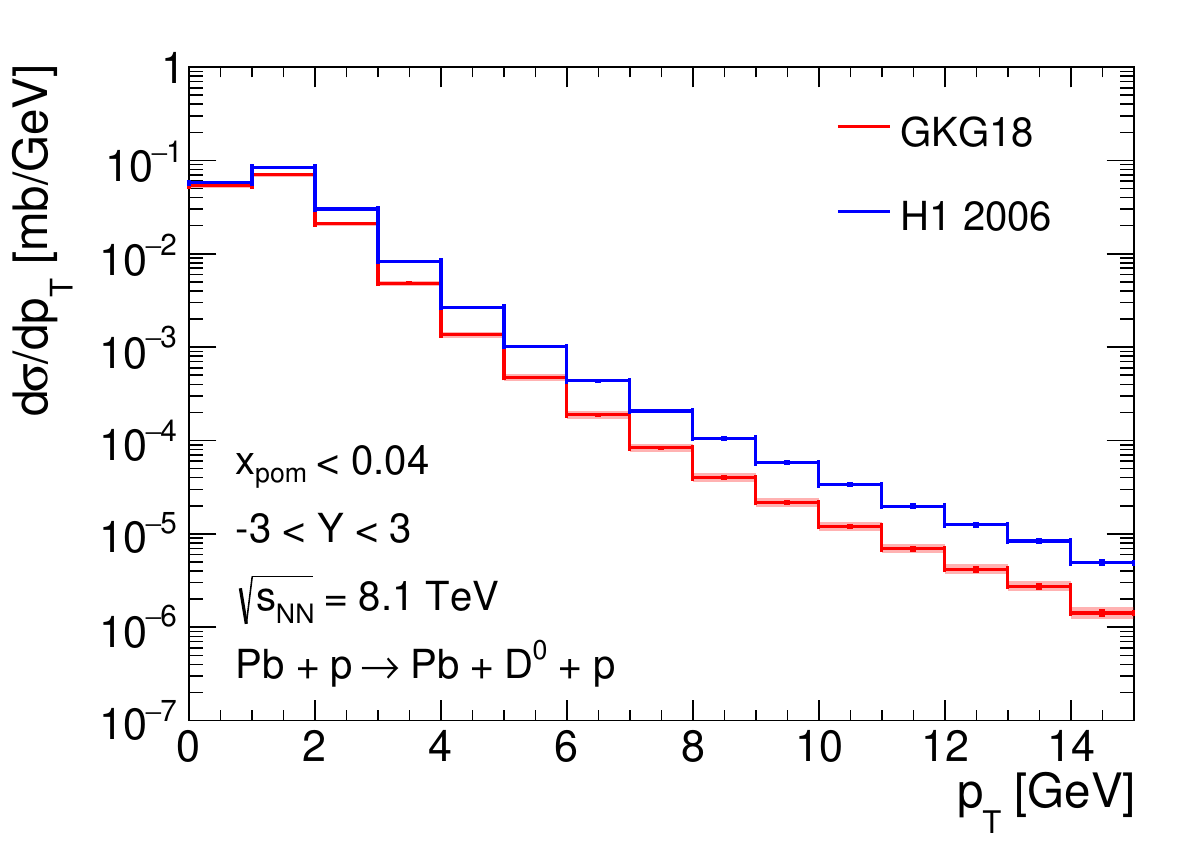}
    \includegraphics[width=0.49\linewidth]{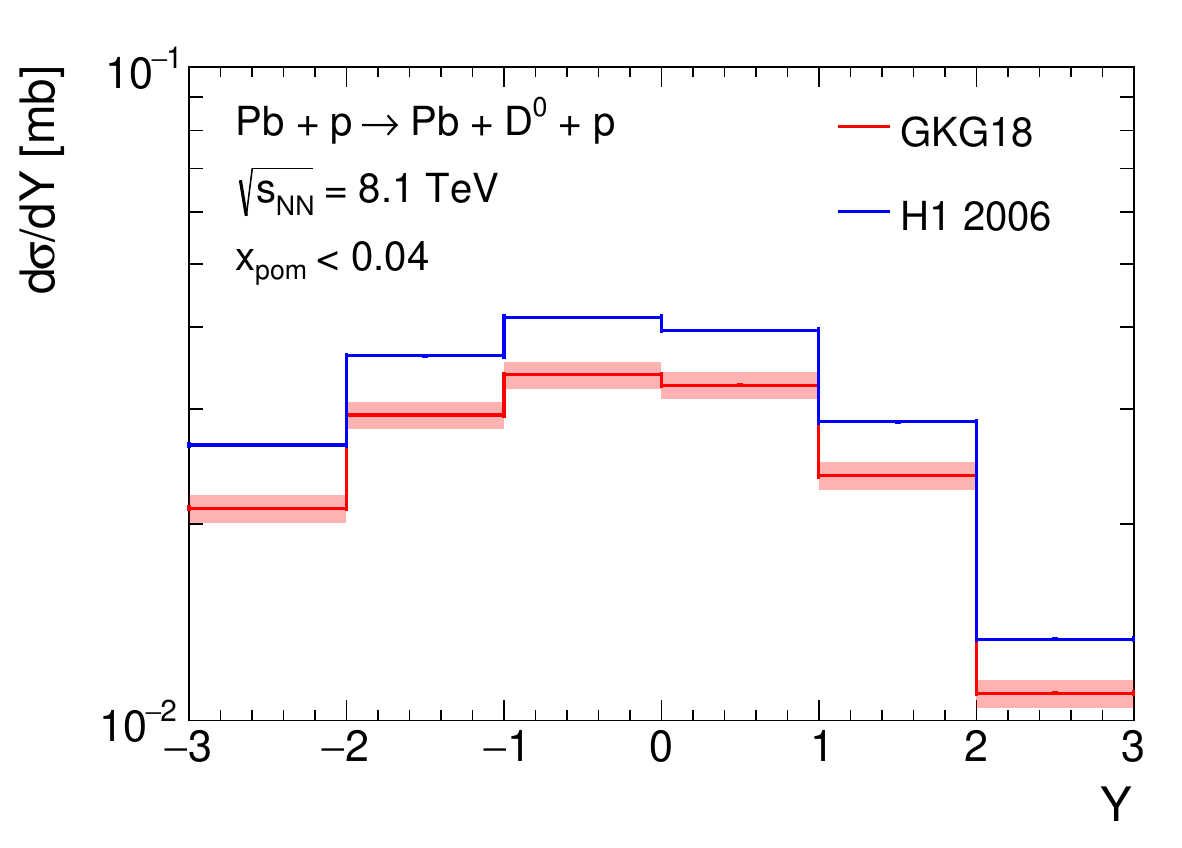}
    \caption{Predictions for the transverse momentum (left panel) and rapidity (right panel) distributions for the diffractive $D^0$ photoproduction   in ultraperipheral $pPb$ collisions, derived using the \textsc{Pythia}~8 framework and assuming distinct diffractive PDFs for the proton.}
    \label{fig:difUPC_distpty}
\end{figure}

\begin{figure}
    \centering
    \includegraphics[width=0.49\linewidth]{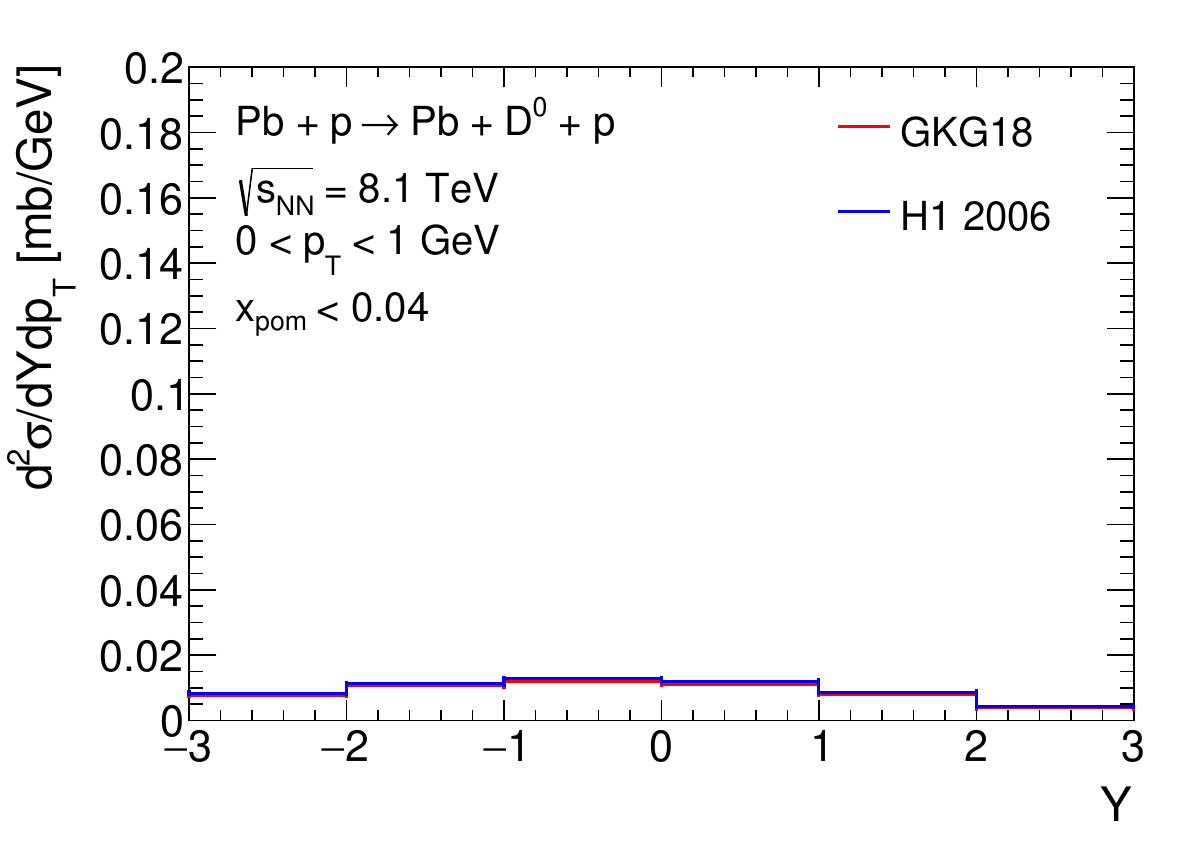}
    \includegraphics[width=0.49\linewidth]{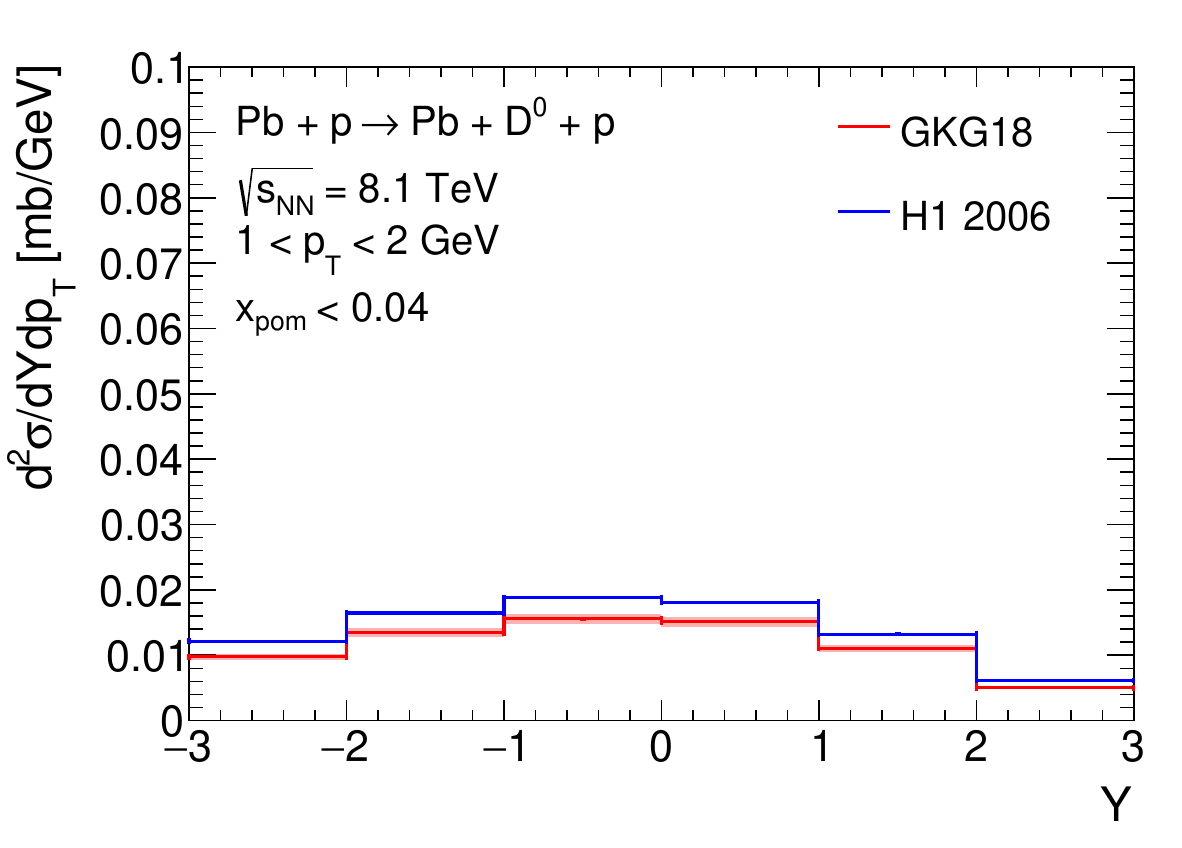}
    \includegraphics[width=0.49\linewidth]{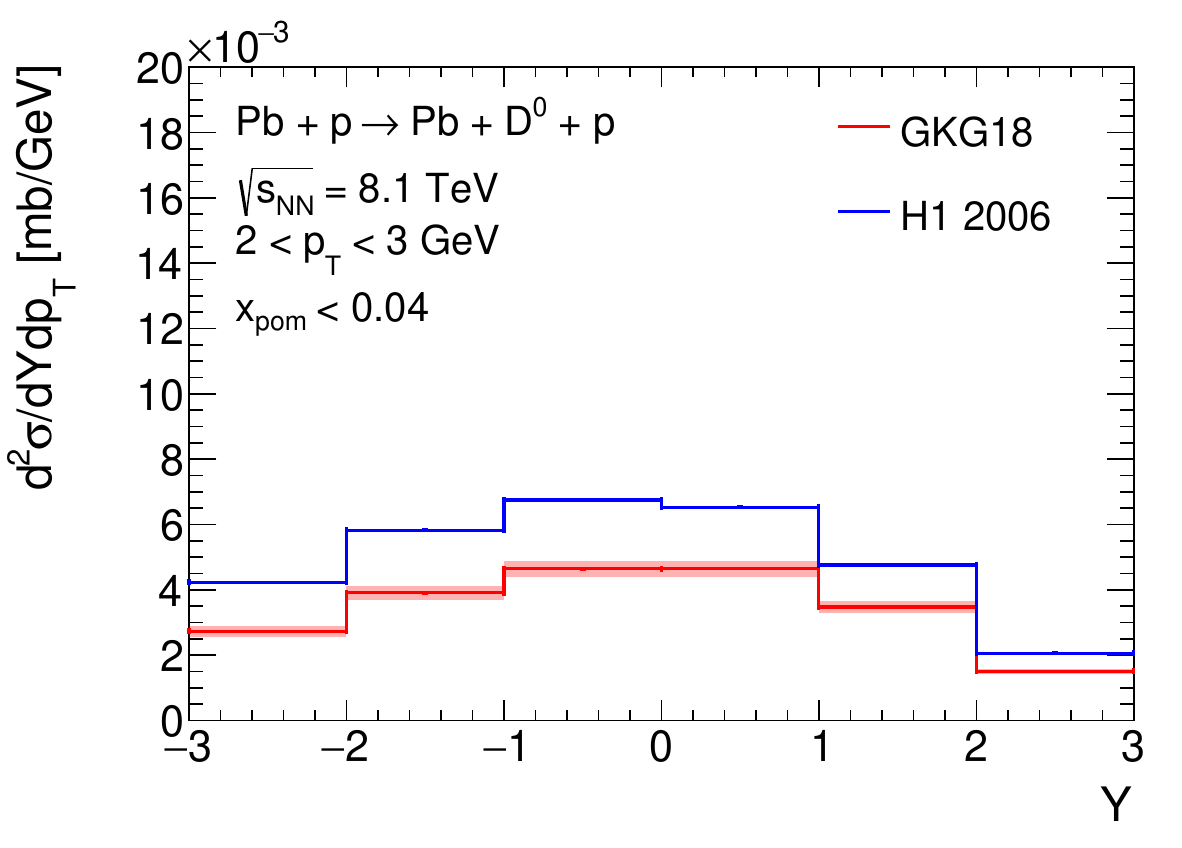}
    \includegraphics[width=0.49\linewidth]{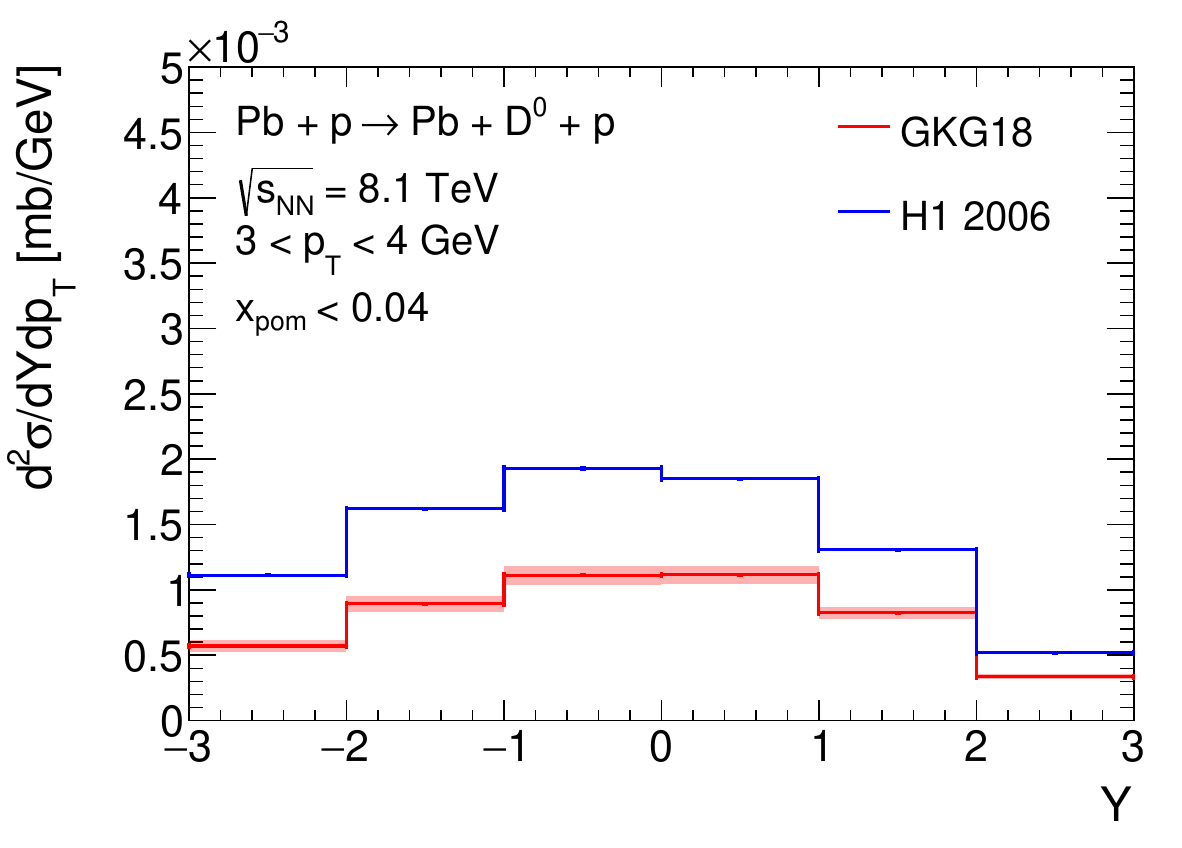}
    \includegraphics[width=0.49\linewidth]{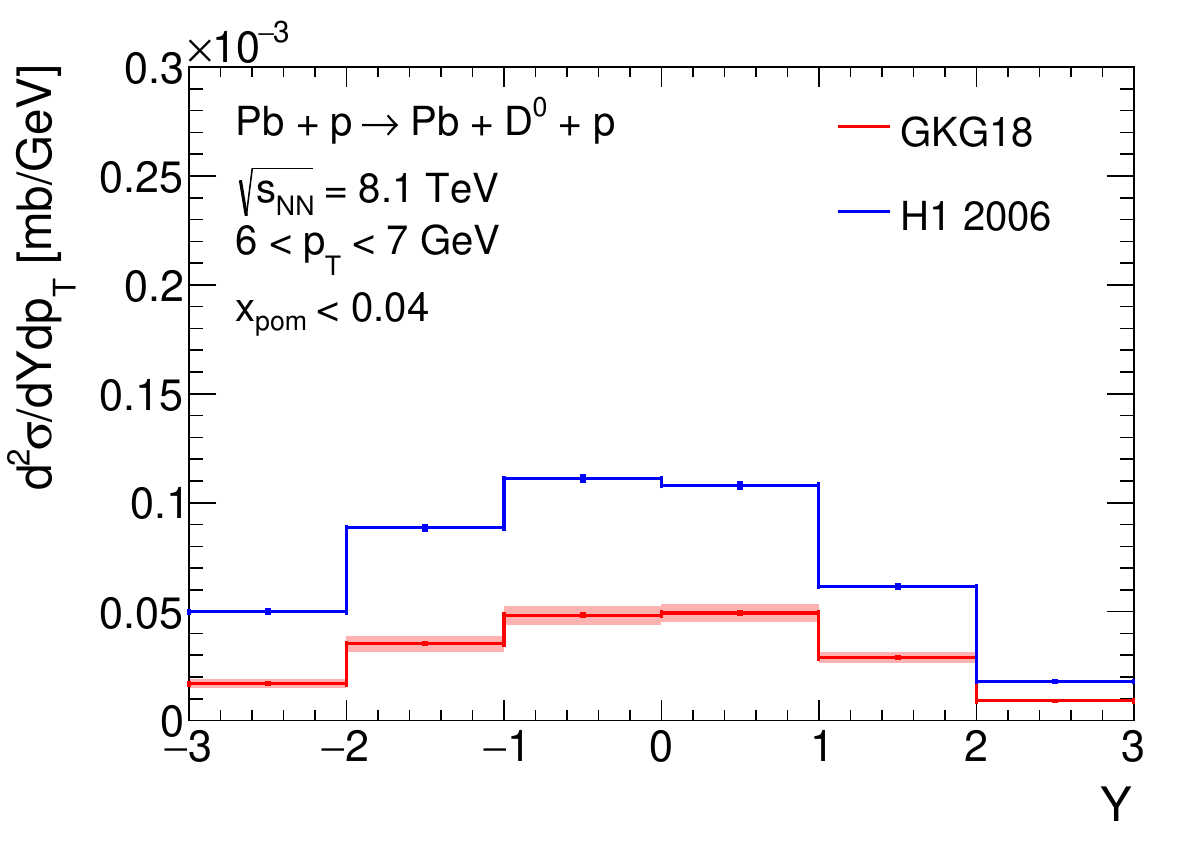}
    \includegraphics[width=0.49\linewidth]{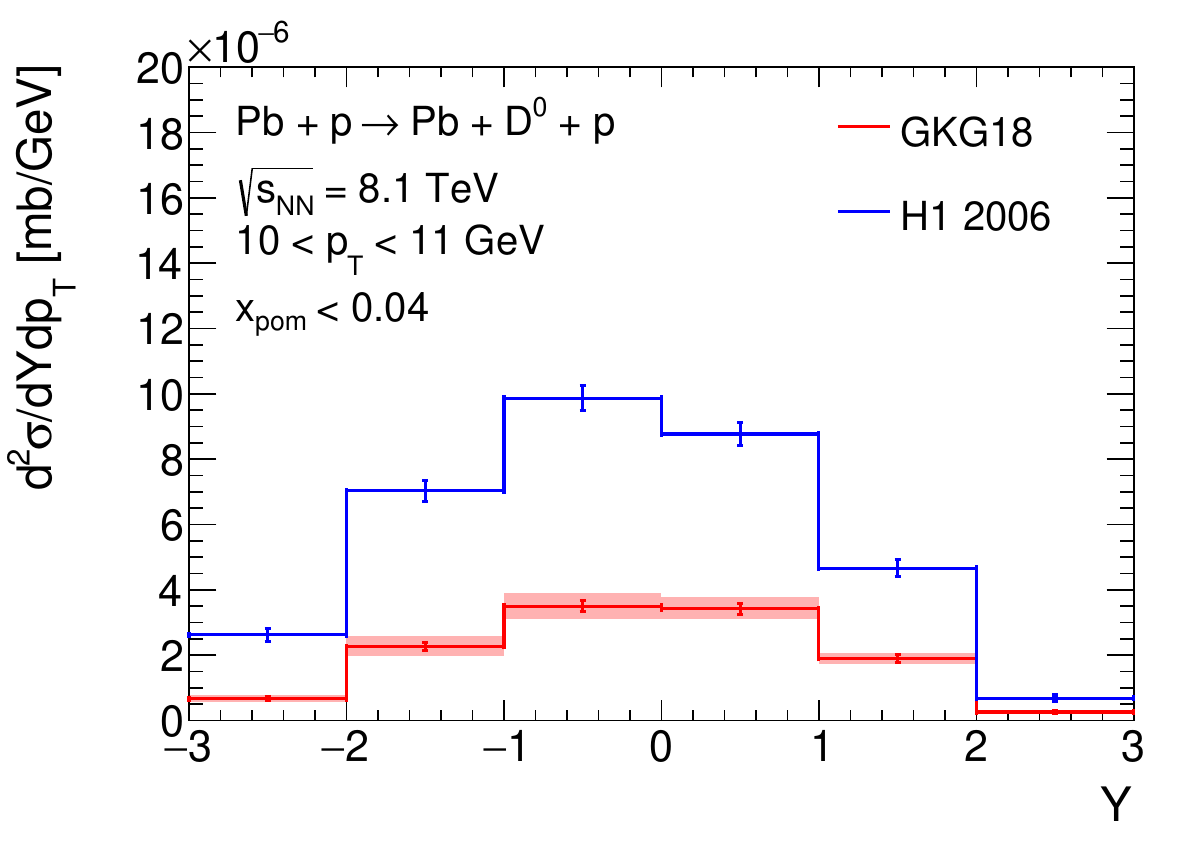}
    \caption{Double-differential cross section for diffractive $D^0$ photoproduction in ultraperipheral $pPb$ collisions, shown as a function of
rapidity. The six panels correspond to the following $p_T$ bins: (0, 1), (1, 2), (2, 3), (3, 4), (6, 7), and (10, 11) GeV.}
    \label{fig:difUPC_ptbins}
\end{figure}

Figure~\ref{fig:difUPC_distpty} shows our predictions for diffractive $D^0$ photoproduction in ultraperipheral $pPb$ collisions. The transverse-momentum spectrum (left) is similar for GKG18 and H12006 at low $p_T$, but the predictions progressively diverge with increasing $p_T$, with GKG18 yielding smaller cross sections. For the rapidity distribution (right), H12006 gives a larger overall normalization in the rapidity range shown. In Fig.~\ref{fig:difUPC_ptbins} we present the rapidity dependence of the double-differential cross section in several $p_T$ bins. The binning and the scales are motivated by Fig.~13 of Ref.~\cite{Guzey:2026lfb}, where the same observable was evaluated using a different theoretical approach and a distinct dPDF. However, the GKG18 results are similar to those of Ref.~\cite{Guzey:2026lfb}. Moreover, the difference between the GKG18 and H12006 predictions grows with $p_T$, consistent with Fig.~\ref{fig:difUPC_distpty}.

\begin{figure}
    \centering
    \includegraphics[width=0.49\linewidth]{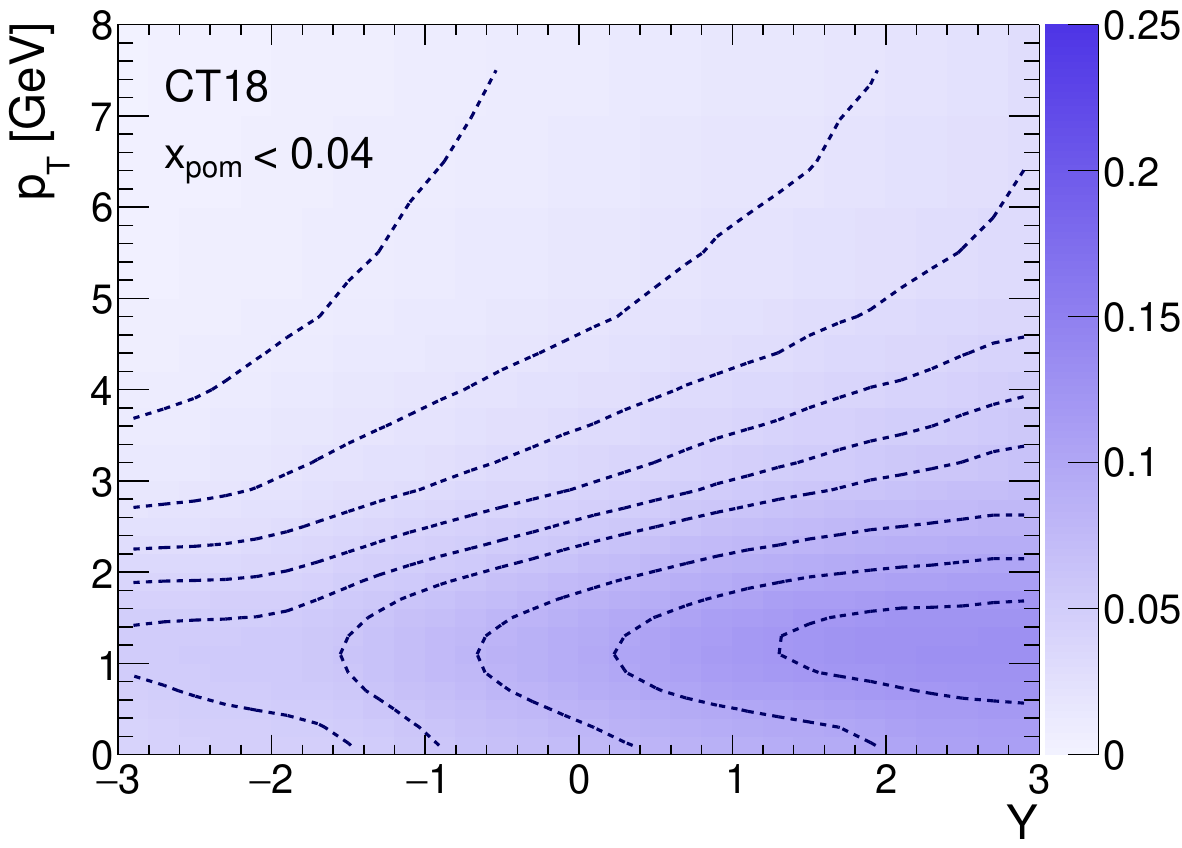}
    \includegraphics[width=0.49\linewidth]{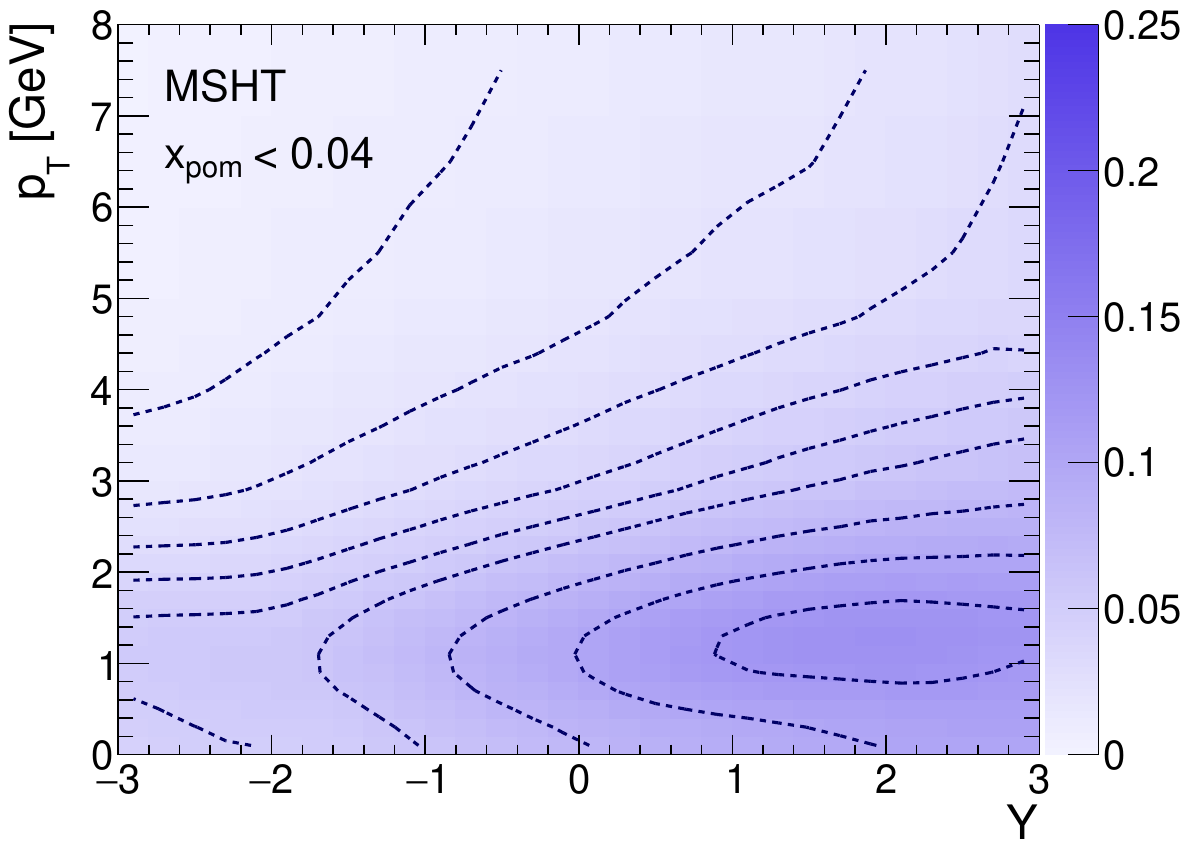}
    \includegraphics[width=0.49\linewidth]{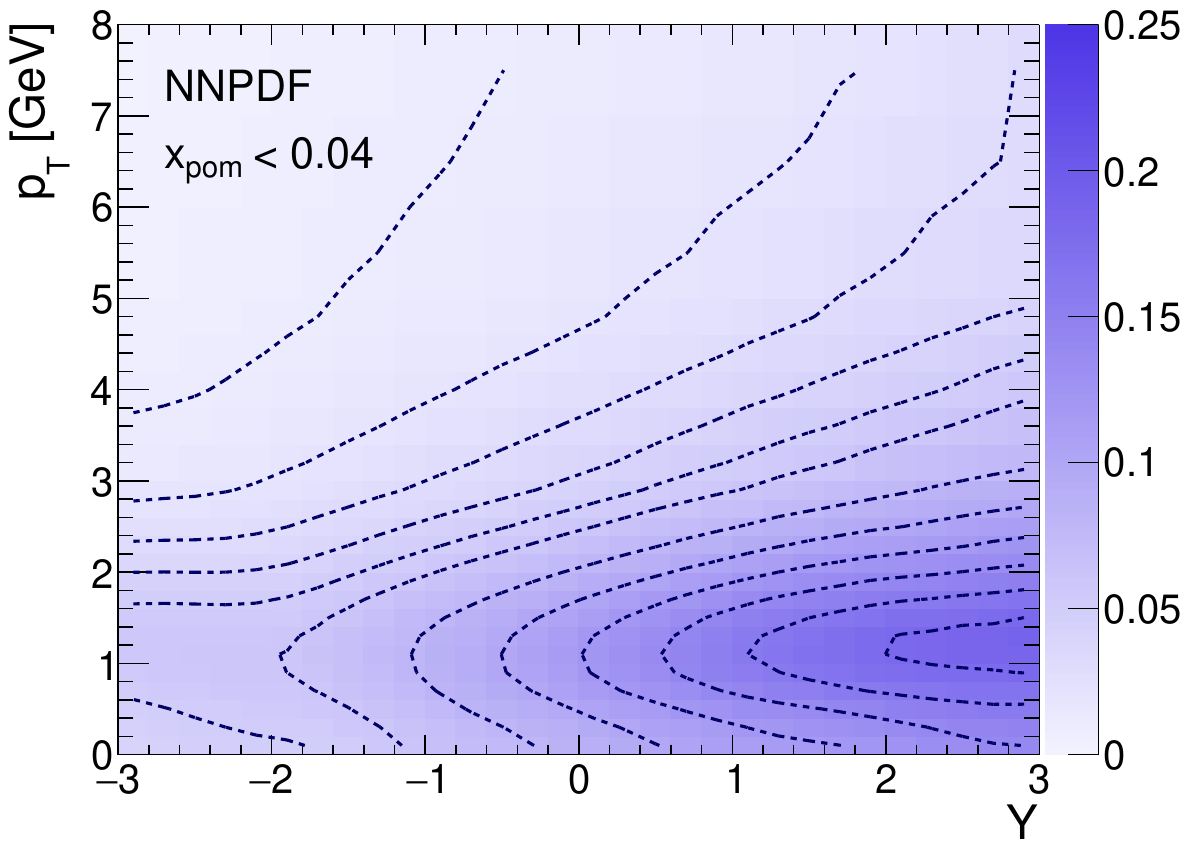}
    \includegraphics[width=0.49\linewidth]{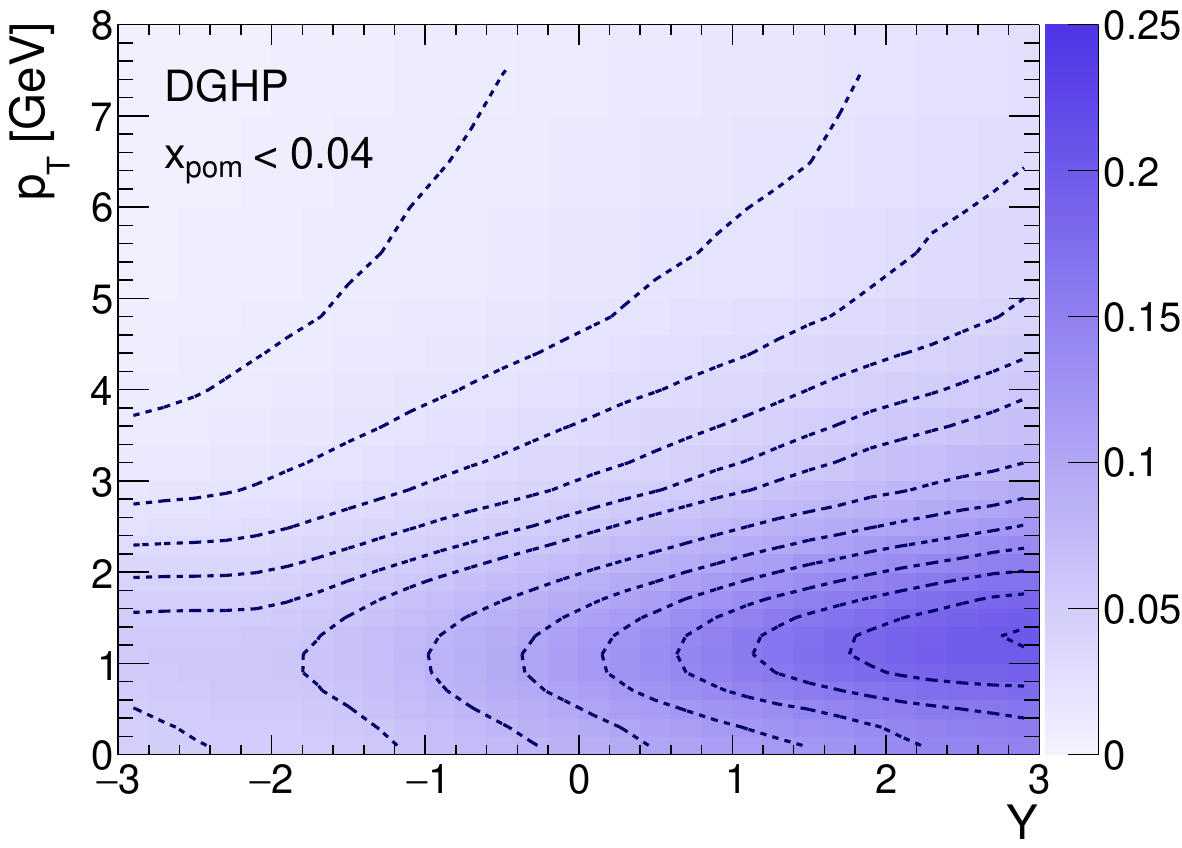}
    \caption{Ratio of the diffractive-to-inclusive $D^0$ photoproduction cross sections in  ultraperipheral $pPb$ collisions $(Y, p_T)$ plane for different proton PDFs. The diffractive
component is integrated over $x_{\pom}$ up to  0.04 and has been calculated using the GKG18 dPDF.}
    \label{fig:ratio}
\end{figure}

In order to quantify the magnitude of the diffractive contribution, we present in Fig.
Fig.~\ref{fig:ratio} the ratio between the diffractive and the inclusive cross - sections in the $(Y, p_T)$ plane for different proton PDFs. For this calculation, the diffractive
component was integrated over $x_{\pom}$ up to  0.04  using the GKG18 dPDF.  In agreement with the results obtained in Ref.~\cite{Guzey:2026lfb}, the diffractive contribution increases for smaller values of $p_T$ and larger rapidities. Our results indicate that the ratio is sensitive to the proton PDF, reaching values larger than 15\% for $p_T \approx 1$ GeV and $Y \ge 2$.

\begin{figure}
    \centering
    \includegraphics[width=0.49\linewidth]{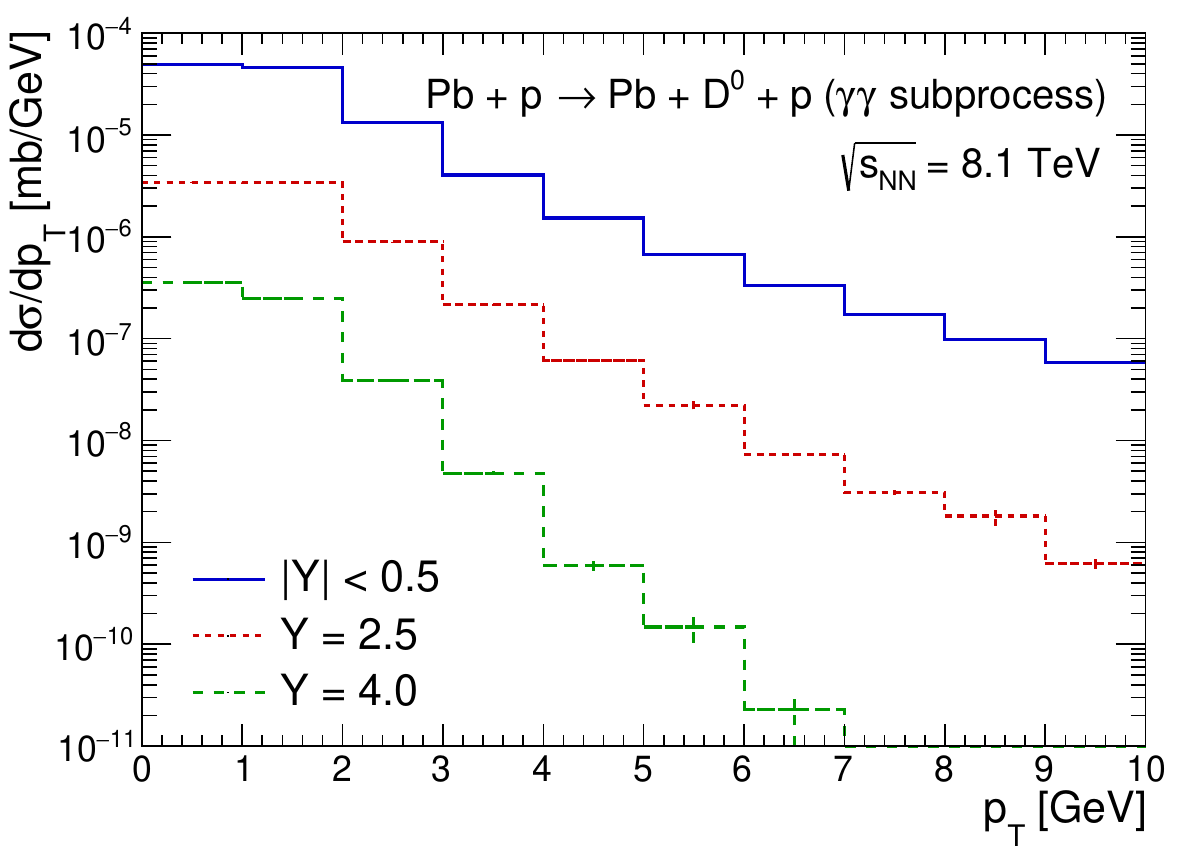}
    \includegraphics[width=0.49\linewidth]{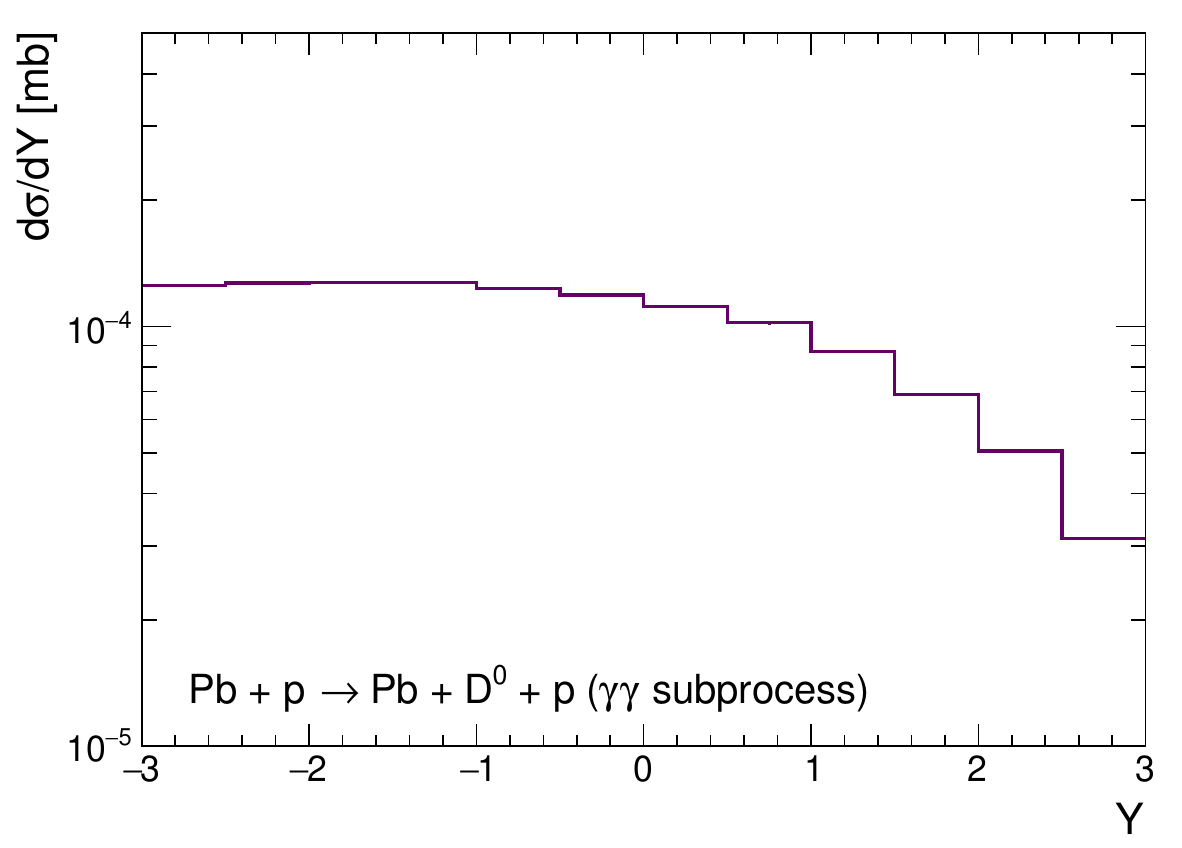}
    \caption{Predictions for the transverse momentum (left panel) and rapidity (right panel) distributions associated with the   $D^0$ meson production via $\gamma\gamma$ interactions in ultraperipheral $pPb$ collisions. }
    \label{fig:gammagamma}
\end{figure}

Finally, Fig. ~\ref{fig:gammagamma} presents our predictions for $D^0$ meson production via $\gamma\gamma$ interactions in ultraperipheral $pPb$ collisions. Similarly to the diffractive case, the final state is characterized by two intact hadrons and two rapidity gaps. While this channel could, in principle, act as a background to diffractive event separation, our results indicate that the two-photon cross-section is nearly two orders of magnitude smaller than the diffractive one. Therefore, this contribution is negligible for $D^0$ meson production in ultraperipheral $pPb$ collisions.

\section{Summary}
\label{sec:summary}

In recent years, it has become feasible to probe inclusive photon-induced processes in ultraperipheral hadronic collisions. In particular, the recent measurement of inclusive $D^0$ photoproduction in ultraperipheral $PbPb$ collisions has motivated improvements in the theoretical description of heavy quark production in photon-nucleus interactions. Moreover, a dedicated analysis of the diffractive $D^0$ photoproduction in these collisions has been proposed as a way to improve our understanding of diffractive interactions and to constrain the nuclear dPDFs.
 These studies also suggest that analyzing this final state in ultraperipheral $pPb$ collisions offers an important alternative for discriminating between competing formalisms and constraining their core assumptions, while avoiding the complications associated with nuclear effects and the treatment of electromagnetic dissociation of the colliding ions.  

Motivated by these earlier studies, this paper presents a systematic analysis of inclusive and diffractive $D$-meson photoproduction in $ep$ and ultraperipheral $pPb$ collisions within the \textsc{Pythia}~8 framework. We compare predictions obtained with different proton and Pomeron PDFs with H1 measurements of inclusive and diffractive $D^*$ photoproduction at HERA, and find that \textsc{Pythia}~8 provides a satisfactory description. This agreement supports extending the framework to $D$-meson photoproduction in ultraperipheral collisions. We presented predictions for the rapidity and transverse-momentum distributions and estimate the diffractive-to-inclusive $D^0$ photoproduction ratio. Our results indicate that the diffractive contribution exceeds 15\% at low $p_T$ and forward rapidities, suggesting that this process could be accessible in future LHC runs. They also indicate that measurements of inclusive and diffractive $D^0$ photoproduction in ultraperipheral $pPb$ collisions can help constrain the proton and Pomeron PDFs.

The results presented in this paper provide a baseline for analyzing nuclear effects in PDFs, which can be further explored by studying inclusive and diffractive $D^0$ photoproduction in ultraperipheral $PbPb$ collisions using the \textsc{Pythia}~8 framework.

\begin{acknowledgments}
The authors acknowledge helpful discussions with Ilkka Helenius and  Souvik Adhya regarding the implementation of the photoproduction process in \textsc{Pythia}~8.   V.P.G. and L. S.  were partially supported by CNPq, CAPES (Finance Code 001), FAPERGS and INCT-FNA (Process No. 408419/2024-5). The work of W.S. was partially supported by the Polish National Science Center Grant No. UMO-2023/49/B/ST2/03665.
\end{acknowledgments}

\bibliographystyle{unsrt}

\end{document}